%% file: tapsse.tex
\pdfoutput=1
\documentclass[sigconf]{acmart}
\setcopyright{acmcopyright}
\acmDOI{10.1145/3377811.3380428}
\acmISBN{978-1-4503-7121-6/20/05}
\acmYear{2020}
\copyrightyear{2020}
\acmPrice{15.00}
\acmConference[ICSE '20]{42nd International Conference on Software Engineering}{May 23--29, 2020}{Seoul, Republic of Korea}
\acmBooktitle{42nd International Conference on Software Engineering (ICSE '20), May 23--29, 2020, Seoul, Republic of Korea}

\usepackage[linesnumbered,ruled]{algorithm2e}
\makeatletter
\renewcommand{\@algocf@capt@plain}{above}
\makeatother
\usepackage{algorithmicx}
\usepackage{algpseudocode}    
\usepackage{amsmath}
\usepackage{amssymb}
\usepackage{balance}
\usepackage{bbding}
\usepackage{booktabs}
\usepackage{calc}
\usepackage{circuitikz}
\usepackage{color}            
\usepackage{etex}
\usepackage{filecontents}
\usepackage{graphicx}
\usepackage[customcolors]{hf-tikz}
\usepackage{listings}
\usepackage{mdwlist}          
\usepackage{microtype}        
\usepackage{multirow}
\usepackage{paralist} 
\usepackage{pgfplots}
\usepackage{pifont}
\usepackage{pst-circ}
\usepackage{pstricks}
\usepackage{relsize}
\usepackage{setspace}
\usepackage{stmaryrd}
\usepackage{subfigure}
\usepackage[T1]{fontenc}      
\usepackage{tikz}
\usepackage[utf8]{inputenc}   
\usepackage[USenglish]{babel} 
\usetikzlibrary{matrix,arrows,positioning,shapes,backgrounds,fit}
\pgfplotsset{compat=1.3}
\usepackage{wasysym}
\usepackage{xcolor}

\newcommand\ignore[1]{}

\newcommand{\textcode}[1]{\texttt{#1}}

\newcommand*\circled[1]{\tikz[baseline=(char.base)]{
				\node[shape=circle,draw,inner sep=1pt] (char) {\scriptsize{#1}};}}

\newcommand{\pcon}{\mathit{pc}}

\newcommand{\outline}[1]{}

\newcommand{\SpecuSym}{\textsc{SpecuSym} }
\newcommand{\SymExec}{\textsf{BaseSymExec} }

\let\oldnl\nl
\newcommand{\nonl}{\renewcommand{\nl}{\let\nl\oldnl}}

\newcommand{\prog}{\mathcal{P}}

\definecolor{darkblue}{rgb}{0, 0.125, 0.576}
\definecolor{dkgreen}{rgb}{0,0.6,0}
\definecolor{gray}{rgb}{0.5,0.5,0.5}
\definecolor{mauve}{rgb}{0.58,0,0.82}

\newcommand\sgnote[1]{\textcolor{blue}{{\textbf{Daniel Says: #1}}}}

\newcommand{\submissioncomment}[1]{  }

\newtheorem{pro}{Property}

\definecolor{salmon}{RGB}{255,191,191}
\tikzset{
  woval/.style={minglingh
    draw
    , line width=1pt
    , anchor=center
    , text centered
    , rounded corners
  },
}
\tikzset{
  hoval/.style={
    drawhttps://www.overleaf.com/project/5e2b89ed125b5500011bf213
    , line width=1pt
    , fill=salmon
    , anchor=center
    , text centered
    , rounded corners
  },
}



\begin{document}

\setlength{\abovedisplayskip}{3pt}
\setlength{\belowdisplayskip}{3pt}

\title{\textsc{SpecuSym}: Speculative Symbolic Execution for Cache Timing Leak Detection}

\author{Shengjian Guo}
\authornote{Both authors contributed equally to this research. Yueqi Chen worked on this 
	project while he interned at Baidu USA.}
	\affiliation{%
		\institution{Baidu Security}
}
\email{sjguo@baidu.com}

\author{Yueqi Chen}
\authornotemark[1]
\affiliation{%
  \institution{Penn State University}
}
\email{yxc431@ist.psu.edu}

\author{Peng Li, Yueqiang Cheng}
\affiliation{%
  \institution{Baidu Security}
}
\email{{lipeng28,chengyueqiang}@baidu.com}

\author{Huibo Wang}
\affiliation{%
  \institution{Baidu Security}
}
\email{wanghuibo01@baidu.com}

\author{Meng Wu}
\affiliation{%
  \institution{Ant Financial Services Group}
}
\email{bode.wm@antfin.com}

\author{Zhiqiang Zuo}
\affiliation{%
  \institution{State Key Lab. for Novel Software Technology, Nanjing University}
}
\email{zqzuo@nju.edu.cn}


\begin{abstract}
  CPU cache is a limited but crucial storage component in modern processors, 
	whereas	the	cache timing side-channel may inadvertently leak information 
	through the physically measurable timing variance. Speculative execution, 
	an essential processor optimization, and a source of such variances, can 
	cause severe detriment on deliberate branch mispredictions.
  Despite static analysis could qualitatively verify the timing-leakage-free 
	property under speculative execution, it is incapable of producing endorsements 
	including inputs and speculated flows to diagnose leaks in depth. 
  This work proposes a new symbolic execution based method, \textsc{SpecuSym}, 
	for precisely detecting cache timing leaks introduced by speculative execution. 
	Given a program (leakage-free in non-speculative execution), \textsc{SpecuSym} 
	systematically explores the program state space, models speculative behavior at 
	conditional branches, and accumulates the	cache side effects along with subsequent 
	path explorations. 
	During the dynamic execution, \textsc{SpecuSym} constructs leak predicates for 
	memory visits according to the specified cache model and conducts a 
	constraint-solving based cache behavior analysis to inspect the new cache behaviors.
  We have implemented \textsc{SpecuSym} atop KLEE and evaluated it against 15 
	open-source benchmarks. Experimental results show that \SpecuSym successfully 
	detected from 2 to 61 leaks in 6 programs under 3 different cache settings 
	and identified false positives in 2 programs reported by	recent work.
\end{abstract}



\maketitle

\renewcommand{\shortauthors}{Guo et al.}

\section{Introduction}
\label{sec:intro}

CPU cache is a limited but crucial storage area on modern processor chips. 
It primarily relieves the speed disparity between the rapid processors and 
the slow main memory, by buffering recently used data for faster reuse. Cache 
timing side-channel attacks~\cite{Kocher96,DhemKLMQW98} leverage the distinct 
cache physical symptoms, i.e., the cache visiting latencies of various program 
executions, to penetrate the confidentiality of the victims. 
On exploiting the vulnerable software implementations, adversaries can extract 
the application secrets
\cite{OsvikST06,TromerOS10,GullaschBK11,CGM16}, infer the neural network 
structure~\cite{YanFT18,HuLDLXJDLSX18,HongDKLRKDD18,DudduSRB18}, or even dump 
the kernel data
\cite{HundWH13,LippSGPHFHMKGYH18,KocherGGHHLMPSY19,WeisseVMGKPSSWY18}.

A timing side-channel generally serves as the intermediate carrier through
which private data could inadvertently disclose to observers who can elaborately 
measure the timing information of certain operations. One particular instance 
is the cache timing side-channel, which leaks data by the variance of the 
cache visiting latency. State-of-the-art program repair method~\cite{WuGSW18} 
mitigates cache timing leaks by enforcing constant execution time for all 
secret relevant operations. However, this strong mitigation may still get 
compromised by the thread-level concurrency~\cite{GuoWW18} or the 
instruction-level parallelism like \textit{speculative execution}~\cite{kimuraKT1996}.

\textit{Speculative execution}~\cite{kimuraKT1996} is a microarchitectural 
optimization in modern processors. It primarily increases the CPU instruction 
pipeline throughput by beforehand scheduling instructions under predicted 
branches, which prevents control hazards from stalling the pipeline. Despite 
its essential importance, the cache side effects caused by prediction errors 
could engender severe detriment through the cache timing side-channel
\cite{KocherGGHHLMPSY19,BulckMWGKPSWYS18,WeisseVMGKPSSWY18,IslamMBKGES19}.

Program analysis for speculative execution is by no means a new research domain. 
Previous efforts mainly researched safe and efficient execution
\cite{ChenLDHY04,PrabhuRV10,GuarnieriKMRS19}, worst-case execution time estimation
\cite{LiMR03,LiMR05}, concurrency bug prediction~\cite{ChenWYS09,LiELS05}, and 
Spectre vulnerability detection~\cite{GuarnieriKMRS19,OleksenkoTSF19,WangCBMR19}. 
Wu et al.~\cite{WuW19} recently proposed a dedicated static  analysis of 
timing-leakage-free property under speculative execution. However, this abstract 
interpretation based method~\cite{WuW19} qualitatively answers the \textit{yes} or 
\textit{no} question --- it is incapable of generating input and speculative flows 
to diagnose leaks in depth. Moreover, the over-approximation nature inevitably 
results in false positives, which desires a more precise method.

\begin{figure}
  \centering
  \scalebox{1.0}{\input{fig_flow.tex}}
  \caption{Overall flow of \textsc{SpecuSym}.}
  \label{fig:overall_flow}
	\vspace{-2ex}
\end{figure}
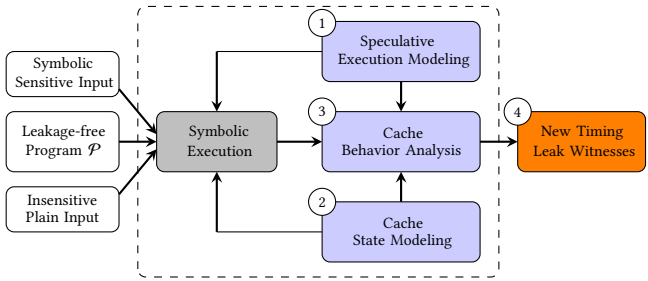

To this end, we propose a new symbolic execution based method, \textsc{SpecuSym}, 
for detecting cache timing leaks caused by speculative 
execution. Figure~\ref{fig:overall_flow} displays the overall flow of 
\textsc{SpecuSym}. Given a program $\prog$, which is timing-leakage-free in 
non-speculative execution, the sensitive input presented in symbol, and the 
insensitive input, \textsc{SpecuSym} leverages symbolic execution to explore 
$\prog$'s state space systematically. Meanwhile, it models speculative execution 
at conditional branches (cf. \circled{1}) and accumulates cache side effects
along with subsequent executions (cf. \circled{2}). Based on a cache model, 
\textsc{SpecuSym} constructs leak predicates for memory visits and conducts a 
constraint-solving based cache behavior analysis (cf. \circled{3}) to generate 
the leak witnesses (cf. \circled{4}).

Our new method has three significant challenges. The first challenge comes from 
the modeling of speculative behaviors. Classic symbolic executors
\cite{CadarDE08,PasareanuR10} 
neither support speculative execution nor are cache-aware since they primarily 
concentrate on the functional correctness rather than reasoning the implicit 
program properties. 
The second challenge derives from the cache state maintenance. Due to the symbolic 
nature, a symbolic memory address may correspond to multiple concrete addresses. 
Updating the cache status after each memory operation unquestionably leads to an
explosive number of different cache states. 
The last challenge stems from the analysis cost. Processors may trigger multiple 
branch mispredictions during program execution. Indiscriminately covering all 
possibilities introduces not only tremendous constraint solving overhead but also 
many unnecessary cases.

To overcome the first challenge, we design a new modeling algorithm in symbolic 
execution, which satisfies both feasibility and high-fidelity. In essence, it 
utilizes the stateful exploration to mimic the speculative behaviors and isolates 
memory changes in auxiliary states from the normal symbolic states.
To tackle the second challenge, we develop a lazy modeling strategy that tracks 
memory accesses and lazily reasons about cache effects rather than maintaining a 
complete set of cache states on-the-fly. 
To address the last challenge, we filter the branches that are unlikely to cause 
harmful speculative execution. Also, we develop several optimizations to shrink 
the constraint size for solving cost reduction.

\ignore{
Also, We develop a merging schema between the mimicked states and regular symbolic 
states to accumulate the cache side effects.

In general, we decompose the precise but lengthy constraint into smaller chunks 
without losing correctness, as well as utilizing executor kernel characteristics 
for faster computation.

However, the side effects caused by speculative execution are normally undetectable 
under standard symbolic execution. To overcome this problem, we introduce the 
$\mathit{speculative~modeling}$ into symbolic execution. Thus, cache side effects 
are visible and cache timing leaks from speculative execution detectable now. 
}

We have implemented \SpecuSym atop KLEE~\cite{CadarDE08} and LLVM~\cite{LattnerA04} 
and evaluated it on 15 benchmarks, which have 8,791 lines of C code in total. Results 
demonstrate that \SpecuSym successfully detects from 2 to 61 leaks in 6 programs under
3 set-associative caches and identifies false positives in 2 programs reported by 
recent work. To summarize, we have made the following contributions:

\begin{itemize}
  \item 
    A novel approach for modeling microarchitectural speculative execution and 
		analyzing the affected cache behaviors in symbolic execution.
  \item 
		The implementation of \textsc{SpecuSym}, which addresses three major challenges
		and supports cache timing leak detection under speculative execution. 
  \item 
    The evaluation of \SpecuSym on 15 open-source benchmarks to demonstrate its 
    effectiveness through revealing from 2 to 61 leaks upon 3 different cache 
		settings.
\end{itemize}

The remainder of this paper is organized as follows. Section~\ref{sec:mtv} 
motivates our work, and Section~\ref{sec:prelim} reviews the background 
knowledge. Section~\ref{sec:specuSE} states the core contributions and
optimizations. Then we conduct experiments in Section~\ref{sec:evaluation} 
and discuss the related work in Section~\ref{sec:related}. Finally, we 
conclude our work in Section~\ref{sec:conclusion}.

\section{Motivation}
\label{sec:mtv}

This section motivates our work with an example. By studying its 
leakage-free cache behavior under non-speculative execution and the new 
leaks caused by speculative execution, we position how \SpecuSym should
facilitate leak detection.

\subsection{Program ${\prog}$ and the Cache Mapping}
\label{sec:leak_example}

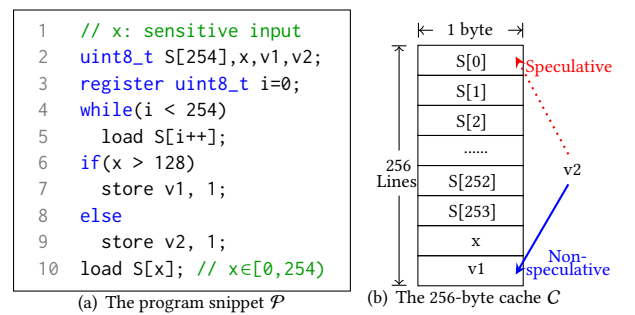
\begin{figure}[htb]
\subfigure[The program snippet $\prog$]{
\label{fig:fig2-a}
\framebox[0.53\linewidth]{     
  \begin{minipage}{0.53\linewidth}
	{\small \tt
  \begin{tabbing} xxx \= xxxx \= \kill
		{\color{gray}{1}} \> \textcolor{dkgreen}{//\texttt{ x: sensitive input}}~~~\\  
		{\color{gray}{2}} \> {\color{blue}{uint8\_t}} S[254],x,v1,v2;\\              
		{\color{gray}{3}} \> {\color{blue}{register uint8\_t}} i=0; \\  
    {\color{gray}{4}} \> {\color{blue}{while}}(i < 254)\\
		{\color{gray}{5}} \> ~~load S[i++]; 		\\
		{\color{gray}{6}} \> {\color{blue}{if}}(x > 128) \\
		{\color{gray}{7}} \> ~~store v1, 1;\\
		{\color{gray}{8}} \> {\color{blue}{else}} \\
		{\color{gray}{9}}\> ~~store v2, 1;\\
    {\color{gray}{10}}\> load S[x]; \textcolor{dkgreen}{//\texttt{ x}$\in$[0,254)}
	\end{tabbing}
  }
	\end{minipage}
}
}
\subfigure[The 256-byte cache $\mathcal{C}$]{
\label{fig:2b}
\begin{minipage}{0.3\linewidth}
	\begin{tikzpicture}[fill=blue!20,font=\footnotesize] 
		\draw[|<-, line width=0.4pt](-0.25, 3.2) -- (-0.25, 1.8);
		\draw[->|, line width=0.4pt](-0.25, 1.2) -- (-0.25, 0);
  	\node[left=4pt, right] at (-0.4, 1.6) {256};
  	\node[left=4pt, right] at (-0.52, 1.4) {Lines};

		\draw (0, 0) rectangle (1.4, 0.4);
  	\node[above=5pt, right] at (0.55, 0.05) {v1};
  	\draw (0, 0.4) rectangle (1.4, 0.8);
		\node[above=5pt, right] at (0.6, 0.4) {x} ;	
  	\draw (0, 0.8) rectangle (1.4, 1.2);
		\node[above=5pt, right] at (0.3, 0.8) {S[253]};
  	\draw (0, 1.2) rectangle (1.4, 1.6);
  	\node[above=5pt, right] at (0.3, 1.2) {S[252]};
  	\draw (0, 1.6) rectangle (1.4, 2.0);
  	\node[above=5pt, right] at (0.5, 1.6) {......};
  	\draw (0, 2.0) rectangle (1.4, 2.4);
  	\node[above=5pt, right] at (0.4, 2.0) {S[2]};
  	\draw(0, 2.4) rectangle (1.4, 2.8);
  	\node[above=5pt, right] at (0.4, 2.4) {S[1]};
  	\draw (0, 2.8) rectangle (1.4, 3.2);
		\node[above=5pt, right] at (0.4, 2.8) {S[0]};

		\draw[|<-, line width=0.4pt](0, 3.4) -- (0.2, 3.4);
		\draw[->|, line width=0.4pt](1.2, 3.4) -- (1.4, 3.4);
  	\node[left=5pt, right] at (0.45, 3.4) {1 byte};

		\node[left=5pt, right] at (2.0, 1.55) {v2};
    \draw [-stealth,dotted,thick,red](2.0, 1.75)--(1.3, 3.05);
    \draw [-stealth,thick,blue](2.0, 1.35)--(1.3, 0.15);
		\node[left=5pt, right,red] at (1.5, 2.9) {Speculative};
		\node[left=5pt, right,blue] at (1.8, 0.4) {Non-};
		\node[left=5pt, right,blue] at (1.5, 0.2) {speculative};

	  \end{tikzpicture}
\end{minipage}
}
\vspace{-3ex}
\caption{Program $\mathcal{P}$ and its cache mapping.}
\label{fig:motiv}
\vspace{-2ex}
\end{figure}

Figure~\ref{fig:fig2-a} shows a program snippet ${\prog}$ whose execution 
time remains stationary in non-speculative execution but varies in terms
of the sensitive input when running under speculative execution.

Listed at line 2, ${\prog}$ has 4 local variables as \texttt{S}, \texttt{x}, 
\texttt{v1}, and \texttt{v2}. Operating these variables, e.g., the implicit 
memory read of \texttt{x} (line 6) and the explicit \emph{store} to \texttt{v1} 
(line 7), may lead to memory access. The remaining variable \texttt{i} (line 3) 
is a register variable that incurs no memory access. Also, variable \texttt{x} 
is the sensitive input, and any form of revealing its value turns to be a leak.

We leverage a fully associative cache $\mathcal{C}$ for the analysis purpose 
of $\prog$, as shown in Figure~\ref{fig:2b}. It is an extreme case of the 
N-way associative cache where the memory address of a variable in $\prog$ may 
map to any cache line of $\mathcal{C}$, subjecting to the line availability 
and the replacement policy. Here we assume $\mathcal{C}$ adopts the \emph{Least
Recently Used} (LRU) policy, which always evicts the least used line once 
$\mathcal{C}$ has been entirely occupied.

Cache $\mathcal{C}$ consists of 256 cache lines, and each line has exactly 
1-byte size. Local variables are mapped to $\mathcal{C}$ according to the 
program execution order and their sizes; e.g., in Figure~\ref{fig:2b} array 
\texttt{S} maps from cache lines \#1 to \#254 because of the array traverse 
in the \emph{while} loop (lines 4-5), and each 1-byte array item successively 
occupies a full cache line. Then variable \texttt{x} associates with line 
\#255. Next, \texttt{v1} fills the last available line (line \#256) if 
\texttt{x}$>$\texttt{128} satisfies; thus, the execution proceeds into the 
\textit{if} flow. Otherwise, the execution takes the \textit{else} branch 
and writes \texttt{v2}, mapping \texttt{v2} to line $\#256$.

\subsection{Leakage-free in Non-speculative Execution}
\label{sec:no_leak}

Running $\prog$ without speculative execution won't leak any information 
about \texttt{x}. This section analyzes the cache behavior in detail.

There are two program paths in $\prog$ because of the \textit{if-else} 
branch. Let's name the one containing the \textit{if} branch as $p_1$ 
and the other one as $p_2$. The only difference between $p_1$ and $p_2$
is the \emph{store} operation, which writes \texttt{v1} on $p_1$ but 
\texttt{v2} on $p_2$. As analyzed in Section~\ref{sec:leak_example}, 
\texttt{v1} would map to the last line of cache $\mathcal{C}$. Similarly, 
without speculative execution, \texttt{v2} also maps to the same line, as 
annotated by the solid arrow in Figure~\ref{fig:2b}. This overlapping 
situation is because usually either $p_1$ or $p_2$ could be taken; 
hence either the \textit{store} to \texttt{v1} or the \textit{store} 
to \texttt{v2} may happen. In both cases, line \#256 is empty before 
the \textit{store} operations. Thereby, \texttt{v1} or \texttt{v2} 
uses this available line.

In view of this fact, we can observe similar cache behaviors on both $p_1$ 
and $p_2$ under non-speculative execution. The first 254 \emph{load} 
operations on array \texttt{S}, the next implicit memory read of \texttt{x},
and the following \textit{store} to \texttt{v1} on $p_1$ or to \texttt{v2} 
on $p_2$, all cause cold misses since cache $\mathcal{C}$ is initially empty. 
Likewise, the last memory \emph{load} on \texttt{S[x]} (line 10) must be a 
cache hit on both paths because \texttt{S} and \texttt{x} have already been 
in cache $\mathcal{C}$. In other words, $\prog$'s cache behavior is 
independent of the sensitive input \texttt{x}. As a result, $\prog$ has no 
cache timing leaks under non-speculative execution.

\subsection{New Leak under Speculative Execution}
\label{sec:specu_leak}

In Section~\ref{sec:no_leak}, only one \textit{store} operation can happen 
in non-speculative execution. However, the situation changes, and a new timing 
leak appears when taking speculative execution into account.

Under speculative execution, the instructions guarded by a branch \emph{br} can 
be scheduled before the execution proceeds into \emph{br} in case CPU predicts 
that \emph{br} is likely to be taken. For example, suppose we first run $\prog$ 
with \texttt{x}$\in$[128,255] several times and flush the cache after each run. 
Afterward, we rerun $\prog$ but setting \texttt{x} to 127. Still, the 
\textit{store} instruction under the \emph{if} branch (line 7) would be executed 
before $\prog$ steers into the \emph{else} branch due to the branch misprediction. 
More importantly, though the CPU performs a rollback to discard the value update 
to \texttt{v1}, \texttt{v1} remains in $\mathcal{C}$ even after the remedy.

Going until line 9, array \texttt{S} and variable \texttt{x} map from lines \#1 
to \#255 and \texttt{v1} still occupies line \#256. At this point, there is no 
empty line available for \texttt{v2}. Following the LRU policy declared in Section
\ref{sec:leak_example}, executing the \textit{store} instruction at line 9 would 
evict the oldest item \texttt{S[0]} from cache $\mathcal{C}$ and map \texttt{v2} 
to the vacated line \#1, as shown by the dotted arrow in Figure~\ref{fig:2b}.

After that, the program execution continues to line 10, reaching the last memory 
\textit{load} of \texttt{S[x]}. Sensitive input \texttt{x} now determines which 
array cell would be visited. And we examine the cache behavior of this \textit{load} 
in terms of two cases \texttt{x=0} and \texttt{x$\neq$0}.

\begin{itemize}
  \item \texttt{x=0}: $\prog$ reads array cell \texttt{S[0]}. Recall that \texttt{x} 
	is still in cache $\mathcal{C}$, but \texttt{S[0]} is no longer in $\mathcal{C}$ 
	due to the replacement by \texttt{v2}. So this memory \textit{load}	causes a 
	conflicting cache miss.
  \item \texttt{x$\neq$0}: Since the whole array \texttt{S} except \texttt{S[0]}
    is in $\mathcal{C}$, the \textit{load} of \texttt{S[x]} must get a cache hit 
	no matter what value \texttt{x} is.
\end{itemize}

Only if \texttt{x=0} there appears one more cache miss on path $p_2$. It is a 
unique situation that enables attackers to learn the value of \texttt{x} due to 
a measurable longer execution time. Note that the cold miss from speculatively 
writing \texttt{v1} also causes an internal latency. However, it is un-observable 
to the external users, and we ignore it safely for analysis purposes.

\subsection{What \SpecuSym Should Provide}
\label{sec:app-scenarios}

The motivating example shows that, though a program may have been carefully 
crafted to avoid cache timing leaks, running it under speculative execution 
could still exhibit new leaks. Since speculative execution is one of the 
fundamental optimizations in modern processors, a systematic analysis that 
detects the subtle leaks would be of vital importance. Specifically, we 
emphasize two requisite abilities in our proposed new method.

First, \SpecuSym should be able to systematically explore the program state 
space to identify execution paths and speculative flows that may cause leaks, 
e.g., $\mathit{p_2}$ and the speculative execution of the \textit{if} branch. 
By contrast, speculation of the \textit{else} branch won't cause any leak along 
with $\mathit{p_1}$.

Second, \SpecuSym should be able to pinpoint the leak sites by a precise cache 
analysis and generate concrete inputs that witness new cache behavior at the 
identified memory visit sites, e.g., the memory \textit{store} instructions to 
\texttt{v2} and the value \textit{zero} of input \texttt{x}.

\section{Preliminaries}
\label{sec:prelim}

This section reviews the preliminary knowledge of symbolic execution, cache timing leak, 
and speculative execution.

\subsection{Symbolic Execution}
\label{sec:se}

Symbolic execution, as a systematic program testing and analysis technique, 
was first introduced in the 1970s~\cite{King76,Clarke76}. In this work, we 
assume that a program $\prog$ consists of a finite set of instructions, and 
$\prog$ defines execution semantics in program paths. Let \textit{inst} be 
an instruction, then interpreting an event $e$:=$(l_b\succ inst\succ l_a)$ 
in symbolic execution stands for the execution of \textit{inst} where $l_b$ 
and $l_a$ denote the locations before and after $inst$, respectively. A 
program execution explores a sequence of events along the program path. 
Close to~\cite{GuoKWYG15,GuoWW18}, we abstract a symbolic event $e$ into 
three categories in terms of the type of \textit{inst} that $e$ contains.

\begin{itemize}
  \item $\psi$-event, which presents a branch instruction. It models the 
    \textit{then} branch by \textcode{assume$(c)$} and the \textit{else} 
    branch by \textcode{assume$(\neg c)$}, respectively. Term $\textit{c}$ 
    is the representative of a conditional predicate expressed in symbolic 
    expression. 

  \item $\chi$-event, which corresponds to a memory \textit{read} instruction 
    of the form $\textit{var}=\textit{load~addr}$, or a memory \textit{write} 
    instruction like $\textit{store~addr, expr}$ where \textit{addr} is the 
    memory address, and \textit{expr} is a symbolic expression.

  \item $\varphi$-event, which represents other types of instructions in the 
    form $\textit{var}:=\textit{expr}$. Here \textit{var} is a variable, and 
    \textit{expr} is a symbolic expression computed from preceding events like 
    arithmetic calculation, bit manipulation, etc.
\end{itemize}

Although symbolic executors always support a rich set of instructions, we use 
the above event types to abstract away the internal implementation details like 
memory allocation, function return, and et al., to center on the high-level flow 
of symbolic execution.

\begin{algorithm}
\caption{Baseline Symbolic Execution.}
\label{alg:baseline}
{\footnotesize
\SetAlgoLined
\setstretch{0.7}
\DontPrintSemicolon
\nonl \textbf{Initially}: The global state container \textsc{Stack} is empty (\textsc{Stack}$\Leftarrow$$\emptyset$).\\
\nonl Start \textbf{\SymExec}($\mathit{s_{ini}}$) on an initial state $\mathit{s_{ini}}$ with $\mathit{in:=\{\lambda,t\}}$.\\
\textbf{\SymExec}(SymbolicState $s$)\\
\Begin{
     $\textsc{Stack}$.push($s$);\;
	 \uIf{$\mathit{s.e}$ is branch event}{
		\For{$\mathit{c \in s.brs}$ {\bf and} $\mathit{s.pc\wedge c}$ is satisfiable} {
				$\mathit{s.pc\leftarrow s.pc\wedge c}$;\\
				\textbf{\SymExec}($\mathit{SubsequentState}$($\mathit{s}$));~~ \textcolor{dkgreen}{// The $\psi$ event}
		}
	}\uElseIf{$\mathit{s.e}$ is memory access event}{
		\textbf{\SymExec}($\mathit{SubsequentState}$($\mathit{s}$));~~ \textcolor{dkgreen}{~~~~~~~// The $\chi$ event}
	}\uElseIf{$\mathit{s.e}$ is other interpretable event}{
		\textbf{\SymExec}($\mathit{SubsequentState}$($\mathit{s}$));~~~~~~~~~ \textcolor{dkgreen}{// The $\varphi$ event}
	}\Else{
		Terminate state $s$ and report error;\;
	}
	$\textsc{Stack}$.pop();\;
}
\BlankLine
\textit{SubsequentState} (SymbolicState \textit{s})\;
\Begin{
  $\mathit{s'} \leftarrow$ symbolically execute $\mathit{e.inst}$ in $\mathit{s}$;\;
  $\mathit{s'.e} \leftarrow$ next available event;\;
  \textbf{return} $s'$;
}
}
\end{algorithm}

Algorithm~\ref{alg:baseline} presents the baseline symbolic execution of a 
program with sensitive input. Unlike prior works~\cite{GuoKWYG15,GuoWW18}, 
both global and local memory accesses are uniformly handled in this algorithm. 
The data input, $\mathit{in}:=\{\lambda,t\}$, determines a program execution 
path consisting of ordered events $\{e_1,\ldots,e_n\}$ where $\mathit{\lambda}$ 
is the sensitive input, e.g., privacy data or cipher keys, and $\mathit{t}$ is 
the insensitive input.

The global container \textsc{Stack} is used for storing symbolic states during 
the dynamic exploration. A symbolic state exhibits the frontier of a path execution. 
We use a tuple $\langle \pcon, \mathit{e}, \mathit{brs}, \Omega, \rangle$ to define 
a state $\mathit{s}$. Symbol $\pcon$ denotes the path condition that leads to 
$\mathit{s}$, where $\mathit{e}$ is the event to execute at $\mathit{s}$. 
$\mathit{brs}$ contains the set of branch predicates if $\mathit{e}$ is a $\psi$ 
event. And $\Omega$ is the symbolic memory which maintains the symbolic memory 
values of the program variables at state $\mathit{s}$.

Initially, the state container \textsc{Stack} is empty, and we start the main 
procedure \SymExec with the initial state $\mathit{s_{ini}}$ on input $\mathit{in}$.
During the recursive execution, \SymExec may split a branch (lines 4-8), perform 
a memory operation (lines 9-10), conduct an internal computation (lines 11-12), 
or terminate a state (line 14), depending on the type of the event to execute.

Note that at the entry of each recursion, \SymExec takes a new symbolic state 
as the input, which is obtained from invoking a secondary procedure 
\textit{SubsequentState}. This procedure inputs the current state \textit{s} 
and outputs a new state $s'$ by symbolically executing the \textit{inst} in 
\textit{s.e}. For brevity, we omit the details of the instruction interpretation, 
which can be found in~\cite{CadarDE08,PasareanuR10}.

\subsection{Cache Timing Leak}
\label{sec:leak}

Memory operations are prone to timing leaks because of the outstanding accessing 
latency between the cache and the main memory. For example, reading data from the 
cache may cost 1-3 processor cycles, whereas loading data from memory could spend 
hundreds of cycles. In this section, we first establish the threat model and then 
formalize the leak definition based on the threat model.

\subsubsection{The Threat Model}
\label{sec:threat}

As exhibited in Section~\ref{sec:specu_leak}, sensitive data involved in memory 
access may get leaked from the timing traffic of a memory write. To capture such 
kind of subtle leak in our analysis, we assume the attackers can perform strong 
external threats.

First, we assume the attackers share the same processor with the victim process. 
Hence they can learn the shared cache states by probe methods. Second, they are 
allowed to request the execution of the victim process. Third, they can observe 
the latency of the interested memory visits in the victim process. Our threat 
model is close to those used in practical attacks like
\cite{OsvikST06,YaromF14,DisselkoenKPT17}, where the attackers can deduce the 
cache block states by measuring the timing traffic of either the victim or the 
attacker process. Moreover, this model also appears in leak detection techniques 
like~\cite{WangWLZW17,DoychevK17,WichelmannMES18,BrotzmanLZTK2018}. Therefore, 
we believe it is a reasonable model for analysis purposes.

\subsubsection{The Leak Definition}

Formally, we abstract a sensitive data related program $\prog$ to be a function 
$\mathit{F_{\prog}(in)\Rightarrow out}$. $\mathit{F_{\prog}}$ processes the data 
input $\mathit{in}:=\{\lambda,t\}$ (cf. Section~\ref{sec:se}), and returns the 
output $\mathit{out}$. For example, assuming $\prog$ is an encryption process, 
$\lambda$ is the private key, and $t$ is the content to encrypt; then, $out$ is 
the ciphertext. Let $T(F_{\prog}(in))$ denotes the execution time of $\prog$ with 
input $\mathit{in}$ under non-speculative execution. Different inputs may explore 
various program paths. However, since we focus on timing leaks introduced by 
speculative execution, here we assume the time of those non-speculative program 
executions remain similar or the same, no matter what the sensitive inputs are, 
which is, 
\begin{multline} 
  \label{lb:no_leak_std}
  ~~~~~\mathit{\forall t,\lambda,\lambda'~.~T(F_{\prog}(\lambda,t)) \simeq 
  T(F_{\prog}(\lambda',t))}~~~~
\end{multline} 
Symbols $\lambda$ and $\lambda'$ denote any two sensitive inputs, and $t$ is still 
the public input. Nevertheless, since in practice, attackers may gain information 
by observing several memory visits or cache lines, we restrict the leak granularity 
at the memory operation level regarding our threat model. Specifically, let $\mathit{e}$ 
be a memory event on executing $\prog$ and the time of interpreting $\mathit{e}$ with 
and without speculative execution be 
$\mathit{T_{s}(\prog_e(in))}$ and $\mathit{T(\prog_e(in))}$, we assume:
\begin{multline} 
  \label{lb:no_leak_se}
  ~~~~~\mathit{\forall t,\lambda,\lambda'~.~T(\prog_e(\lambda,t)) \simeq 
  T(\prog_e(\lambda',t))}~~~~
\end{multline} 
Then the existence of a new cache timing leak under speculative execution can be 
checked by the following formula: 
\begin{multline}
\label{lb:specu_leak}
\mathit{\exists t,\lambda,\lambda'~.~
	\big(
			\lambda \neq \lambda' \wedge T_{s}({\prog_e}(\lambda,t)) 
			\neq T_{s}({\prog_e}(\lambda',t))}
	\big)
	\vee 
	\\  
	T({\prog_e}(\lambda,t)) \neq T_{s}({\prog_e}(\lambda',t)) ~~~
\end{multline}
where a leak appears if (1) two different sensitive inputs can cause significant 
timing differences in executing $\mathit{e}$ under speculative execution, or (2) 
two sensitive inputs can cause significant timing difference in executing $\mathit{e}$ 
with and without speculative execution. In other words, $\prog$ leaks due to 
speculative execution if any pair of $\lambda$ and $\lambda'$ exists. Furthermore, 
we transform formula (\ref{lb:specu_leak}) into a dedicated leak constraint and 
simplify the constraint to a more concise form in Section~\ref{sec:specuSE}. Also, 
as public input plays a minor role in modeling cache timing leaks~\cite{WangBLWZW19}, 
we set insensitive input $t$ to a fixed value to reduce the reasoning cost of formula
(\ref{lb:specu_leak}).

\ignore{
Also, we assume $\prog$ has the same set of instructions on each program path and 
speculative execution only changes the cache state. 
}

\subsection{Speculative Execution}
\label{sec:specu}

The CPU instruction pipeline~\cite{RamamoorthyL77} allows overlapped executions 
of proper instructions where each instruction execution consists of a series of 
micro-stages. This instruction-level parallelism benefits hardware utilization 
since instruction can start execution before the time its prior instructions 
have completed all their stages. However, a pipelined processor may get stalled 
once the program control flow needs to divert, but the destination remains 
unknown (e.g., at a conditional branch). Accordingly, the pipeline has to wait 
until the flow decision gets computed.

To alleviate the cost of such control hazard, processors leverage 
\textit{speculative execution}~\cite{kimuraKT1996} and \textit{branch prediction}
\cite{Mittal19} to reduce the delay that could incur on conditional branch 
instructions. Generally, they predict the execution flow based on the history 
of recently executed branches and schedule instructions under predicted 
branches ahead of jumping into these branches. Specifically, on approaching a 
control hazard, the processors first predict a branch to take. Then, they 
execute the instructions under the selected branch and maintain the temporary 
path state in a dedicated buffer. Finally, they commit the buffered state to 
continue the program flow if they achieved a correct prediction. Otherwise, 
upon an incorrect prediction, they have to discard the temporary state to 
revert the effects of the executed instructions hence avoiding functional errors.

This rollback mechanism, unfortunately, withdraws no affected cache state, which 
raises security risks. As shown in Section~\ref{sec:specu_leak}, variable \texttt{v1} 
maps to cache line \#256 due to the misprediction of the \textit{if} branch; and, 
this cache effect remains unchanged, even control flow directs to the \textit{else} 
branch. As a result, sensitive data \texttt{x} leaks due to the interfered cache 
state. Abstracting away the hardware details, we can model the speculative behavior 
in symbolic execution as a three-phase analysis and evaluate its cache side effects 
with a constraint-solving based approach.

\subsubsection{Misprediction Modeling.}
As aforementioned, on reaching a control hazard, the processors make a prediction to
select a branch for execution. Despite the experience-based hardware realization
in CPUs, we can model this behavior by an auxiliary symbolic state. To be specific, 
before diverging the control flow into the \textit{else} branch, our symbolic executor 
would decide to duplicate a new state from the current symbolic state and schedule
it immediately into the \textit{if} flow, which models the branch prediction of the 
\textit{if} branch. Similarly, we can model the misprediction of the \textit{else} 
branch on demand.

\ignore{
Note that we only model speculative execution at branches where both \textit{if} 
and \textit{else} decisions are feasible in terms of the branch conditions. The 
assumption considers the situation where the hazard is dependent on the computation 
over input and cannot be inferred as a \textit{true} or a \textit{false} value in 
advance. In contrast, if the condition is determined to be \textit{true} or 
\textit{false} at the branch point, e.g., the \textit{while} loop head in Figure
\ref{fig:fig2-a}, then the branch decision is deterministic.

However, different program inputs might go through various program paths to reach 
a same branch point but trigger divergent decisions. Thus, in practice speculative 
executions at these points do exist and our method complies with the fact.
}

\subsubsection{Speculative State Execution.}
Each duplicated state in \textit{misprediction modeling} has the same snapshot 
of its parent state. For clarity, we use the \textit{speculative} state to alias 
this newly forked state. A \textit{speculative} state would be prioritized to 
the front of the state container in the symbolic executor and uninterruptedly 
executes until it meets a predefined threshold, like the size of the Reorder 
Buffer (ROB), the pipeline stage number, the branch depth, etc. Also, a 
\textit{speculative} state runs independently from its parent state; thus, any 
memory updates in the \textit{speculative} state won't taint its parent state 
who is waiting for the \textit{speculative} state returns. For the sake of cache 
analysis, we also maintain the cache data in each symbolic state. A 
\textit{speculative} state inherits such data from the parent state and keeps 
updating the cache during its execution.

\subsubsection{Rollback and Cache Merging.}
Once the \textit{speculative} state reaches the threshold, it has to stop and 
exit the state stack. Right before the termination, it notifies the awaiting 
parent state the finish of the speculation and transfers the latest cache data 
back to its parent. Upon accepting the notification, the parent symbolic state 
merges the received cache information into its cache data, to form the latest 
cache state. After that, the parent state aborts waiting and resumes the regular 
execution. This step models the processor rollback in high fidelity and retains 
the cache status changes from speculation. Since the cache changes have already 
merged into the parent state, terminating the \textit{speculative} state causes 
no further effects.

\ignore{
Our approach is natural to symbolic execution, while state-of-the-art simulation 
method~\cite{OleksenkoTSF19} requires sophisticated instrumentations and special 
instructions to enforce CPU behaviors. More modeling details of our approach would 
appear in Section~\ref{sec:spec_modeling}.
}

\ignore{ 
For the record, speculative execution is *not* just branch prediction. Branch 
prediction is just one type of speculative execution; processors also speculate 
on memory dependencies and reorder loads and stores (called memory disambiguation). 
Branch prediction is *not* inherently out-of-order, since early processors (and 
even some simple ones today) that were pipelined only did in fact do branch 
prediction but never reordered instructions. Superscalar execution, as the authors 
point out Tomasulo's algorithm (btw the citation here is somehow corrupted--please 
fix your bibtex) does in fact reorder instructions, but this is a whole new class 
of reorderings that is not captured in this paper, which leads into (2).
}

It is worth noting that in this work, we constrain the scope of \textit{speculative
execution} to customary \textit{branch prediction} and subsequent out-of-order 
instruction execution. In practice, modern processors may perform other forms of 
speculation, e.g., memory dependence speculation and disambiguation
\cite{Nicolau89,MoshovosS97,ReinmanC98,OnderG99}. However, they are not the main 
focus of this work.

\section{Algorithm}
\label{sec:specuSE}

In this section, we explain the core technical contributions portrayed in Figure
\ref{fig:overall_flow}. Algorithm~\ref{alg:specusym} shows the procedure of 
\textsc{SpecuSym} built upon the baseline algorithm, where the main changes are 
highlighted at the state checkpoint (lines 3-5), the branch point (line 9), and 
the memory access point (lines 13-14).

\begin{algorithm}[t]
\caption{Symbolic Execution in \SpecuSym.}
\label{alg:specusym}
{\footnotesize
\SetAlgoLined
\setstretch{0.7}
\DontPrintSemicolon
\nonl \textbf{Initially}: The global state container \textsc{Stack} is empty (\textsc{Stack}$\Leftarrow$$\emptyset$).;\\
\nonl Start \textbf{\SpecuSym}($\mathit{s_{ini}}$) on an initial symbolic state $\mathit{s_{ini}}$ with $\mathit{in:=\{\lambda,t\}}$.\\
\textbf{\SpecuSym}(SymbolicState $s$)\\
\Begin{
				\If{\textcolor{blue}{$\mathit{s}$ is a speculative state} {\bf{and}} \textcolor{blue}{$\mathit{s}$ reaches the threshold}}{
								\textcolor{blue}{\textbf{return};}\;
					}
				$\textsc{Stack}$.push($s$);\;
        \uIf{$\mathit{s.e}$ is branch event}{
          \For{$\mathit{c \in s.brs}$ {\bf and} $\mathit{s.pc\wedge c}$ is satisfiable} {
            \textcolor{blue}{$\mathit{SpeculativeExplore}$($s$, $c$);}~~~~~~~~~~\textcolor{dkgreen}{// Enter speculative modeling}\;
            ......~~~~~~~~~~~~~~~~~~~~~~~~~~~~~~~~~~~~~~~~~~~~~\textcolor{dkgreen}{// Resume normal execution}\;

          }
        } 
        \uElseIf{$\mathit{s.e}$ is memory access event}{	
          \textcolor{blue}{$\mathit{AnalyzeCache}$($s$);}\;
					\textcolor{blue}{$\mathit{s}.\pi\leftarrow $ update cache state by interpreting $\mathit{s.e}$;}\;
					\textbf{\SpecuSym}($\mathit{SubsequentState}$($\mathit{s}$));\;
        }
        ......\;
}
\BlankLine
$\mathit{SpeculativeExplore}$(SymbolicState $s$, Predicate $c$)\;
\Begin{
  \If{$\mathit{c}$ relies on a memory access}{
    $\mathit{s' \leftarrow }$ duplicate state $\mathit{s}$;~~~~~~~~~~~~~~~~~~~~~~~~~~~~~\textcolor{dkgreen}{// Fork speculative state}\;
    $\mathit{s' \leftarrow redirect~s'~to~the~\neg c~control~flow}$;~~\textcolor{dkgreen}{// Negate branch direction}\;
    \textbf{\SpecuSym}($\mathit{SubsequentState}$($\mathit{s'}$));\;
    $\mathit{s}.\pi \leftarrow \mathit{s'}.\pi$;\;
    Terminate $\mathit{s'}$;\;
  }
}
\BlankLine
$\mathit{AnalyzeCache}$(SymbolicState $s$)\;
\Begin{
    \If{$\mathit{s}$ is a regular symbolic state {\bf{and}} s.e relates to secret input}{
    $\theta\leftarrow$ build the leak constraint for $\mathit{s.e}$;\;
    \If{$\mathit{s.pc}\wedge\theta~\mathit{is~satisfiable}$}{
      Generate witness;\;
    }
  }
}
}
\end{algorithm}

\subsection{Speculative Modeling}
\label{sec:spec_modeling}

This section explains how \SpecuSym models branch misprediction and
speculative execution by introducing an auxiliary speculative state and 
orchestrating it with the normal symbolic state properly.

\subsubsection{Modeling Overview}
On entering the \SpecuSym procedure in Algorithm~\ref{alg:specusym}, we
first check if current state $s$ is a \textit{speculative state} and 
whether $s$ has reached the predefined threshold (line 3). If both 
conditions meet, the recursive symbolic execution on $s$ stops and returns 
immediately (line 4). Note that a state becomes a \textit{speculative} 
state if it is duplicated from a normal symbolic state (line 21) by invoking 
\textit{SpeculativeExplore} at a branch event (line 9).

Lines 18-27 in Algorithm~\ref{alg:specusym} present the modeling procedure 
\textit{SpeculativeExplore}. Generally, at a conditional branch (line 8), 
we call \textit{SpeculativeExplore} (line 9) to start the speculative probe. 
That is, if the branch predicate closely relates to memory visit (line 20), 
e.g., using a variable for the first time, we duplicate a new state $s'$ 
from current state \textit{s} (line 21) with a negated branch direction 
(line 22). This assumption bases on the observation that the memory visiting 
latency at a branch may form a time window for speculative execution to load 
data into the cache. Otherwise, despite a branch misprediction, CPU may spend 
only several cycles on speculative execution before its rollback, which is too 
transient to let speculation affect the cache.

After state duplication, new state ${s'}$ becomes a \textit{speculative} state, 
since it is about to mimic the speculation of the mispredicted branch. We let 
\SpecuSym explore $s'$ (line 23) and assess the effects of the memory accesses 
in $s'$ on the cache (line 15). Details of the assessment present in Section
\ref{sec:modeling}. Once the execution of $s'$ finishes, we use the accumulated 
cache data in $s'$ to update that of $s$ (line 24) and terminate $s'$ (line 25) 
to end its lifecycle. Such design ensures that the speculation of $s'$ only 
contributes to the cache state changes but never affects the memory $\Omega$ of 
its parent symbolic state.

Next, after $\mathit{SpeculativeExplore}$ returns, we resume the normal symbolic
execution of $s$ (line 10), who now has an updated cache. In this way, we have 
constructed a speculative scenario and retained the latest cache changes. Our
modeling method leverages the stateful mechanism of symbolic execution. It not 
only models the speculative behavior but also precisely transfers the cache 
information between symbolic states with controllable flexibility.

\subsubsection{The Motivating Example Revisit}

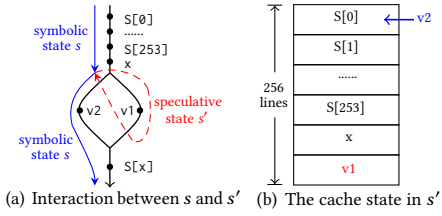
\begin{figure}
\subfigure[Interaction between $s$ and $s'$]{
\label{fig:3a}
\begin{minipage}{0.35\linewidth}
	\begin{tikzpicture}[fill=blue!20,font=\footnotesize] 
		\draw[->, line width=0.5pt, black](0, 0.8) -- (0, 0.275);
		\draw[line width=0.5pt, black](0, 1.8)..controls (-0.55,1.3) .. (0, 0.8);
		\draw[line width=0.5pt, black](0, 1.8)..controls (0.55,1.3) .. (0, 0.8);
		\draw[line width=0.5pt, black](0, 2.7) -- (0, 1.8);

		\filldraw [black] (0, 2.55) circle (1pt);
		\filldraw [black] (0, 2.35) circle (1pt);
		\filldraw [black] (0, 2.15) circle (1pt);
		\filldraw [black] (0, 1.95) circle (1pt);
		\filldraw [black] (-0.4, 1.3) circle (1pt);
		\filldraw [black] (0.4, 1.3) circle (1pt);
		\filldraw [black] (0, 0.55) circle (1pt);

    \node[above=3pt, right, scale=.8] at (0.1, 2.4) {\texttt{S[0]}};
  	\node[above=3pt, right, scale=.8] at (0.1, 2.2) {......};
		\node[above=3pt, right, scale=.8] at (0.1, 2.0) {\texttt{S[253]}};
		\node[above=3pt, right, scale=.8] at (0.1, 1.8) {\texttt{x}};
		\node[above=3pt, right, scale=.8] at (-0.4, 1.2) {\texttt{v2}};
    \node[above=3pt, right, scale=.8] at (0.05, 1.2) {\texttt{v1}};
    \node[above=3pt, right, scale=.8] at (0.1, 0.45) {\texttt{S[x]}};

    \draw [-stealth,thin,blue](-0.2, 2.7)--(-0.2, 1.8);
		\draw [thin,blue](-0.2, 1.8)..controls (-0.3, 1.7) and (-0.6, 1.5) .. (-0.5, 1.1);
		\draw [-stealth,thin,blue](-0.5, 1.1)..controls (-0.3, 0.7) and (-0.2, 0.5) .. (-0.2, 0.3);

		\node[above=3pt, left, scale=.8, blue] at (-0.2, 2.25) {symbolic};
		\node[above=3pt, left, scale=.8, blue] at (-0.3, 2.05) {state $\mathit{s}$};
		\node[above=3pt, left, scale=.8, blue] at (-0.4, 0.8) {symbolic};
		\node[above=3pt, left, scale=.8, blue] at (-0.45, 0.6) {state $\mathit{s}$};

		\draw [densely dashed, thin,red](-0.2, 1.8)..controls (0, 1.9) and (0.4, 1.8) .. (0.5, 1.6);
		\draw [densely dashed, thin,red](0.5, 1.6)..controls (0.6, 1.2) and (0.5, 0.8) .. (0.3, 0.9);
		\draw [-stealth, densely dashed, thin,red](0.3, 0.9)..controls (0.1, 1.2) .. (-0.2, 1.8);

		\node[above=3pt, right, scale=.8, red] at (0.5, 1.3) {speculative};
		\node[above=3pt, right, scale=.8, red] at (0.65, 1.1) {state $\mathit{s'}$};

	  \end{tikzpicture}
\end{minipage}
}
\subfigure[The cache state in $s'$]{
\label{fig:3b}
\begin{minipage}{0.3\linewidth}
	\begin{tikzpicture}[fill=blue!20,font=\footnotesize] 
		\draw[->|, line width=0.4pt](-0.25, 1.7) -- (-0.25, 2.6);
		\draw[->|, line width=0.4pt](-0.25, 1.1) -- (-0.25, 0.2);

  	\node[left=4pt, right, scale=.8] at (-0.35, 1.5) {256};
  	\node[left=4pt, right, scale=.8] at (-0.4, 1.3) {lines};

  	\draw (0, 0.2) rectangle (1.4, 0.6);
		\node[above=5pt, right, scale=.8, red] at (0.55, 0.25) {v1} ;	
  	\draw (0, 0.6) rectangle (1.4, 1.0);
		\node[above=5pt, right, scale=.8] at (0.6, 0.65) {x};
  	\draw (0, 1.0) rectangle (1.4, 1.4);
  	\node[above=5pt, right, scale=.8] at (0.35, 1.05) {S[253]};
  	\draw (0, 1.4) rectangle (1.4, 1.8);
  	\node[above=5pt, right, scale=.8] at (0.5, 1.45) {......};
  	\draw (0, 1.8) rectangle (1.4, 2.2);
  	\node[above=5pt, right, scale=.8] at (0.45, 1.85) {S[1]};
  	\draw(0, 2.2) rectangle (1.4, 2.6);
		\node[above=5pt, right, scale=.8] at (0.45, 2.25) {S[0]};
		\node[above=5pt, right, scale=.8, blue] at (1.55, 2.25) {v2};

		\draw[<-, line width=0.6pt, blue] (1.2, 2.4) -- (1.6, 2.4) ;

	\end{tikzpicture}
\vspace{.3ex}
\end{minipage}
}
\vspace{-3ex}
\caption{The speculative execution modeling of $\prog$.}
\vspace{-3ex}
\label{fig:specu_modeling}
\end{figure}

Figure~\ref{fig:specu_modeling} shows the speculative modeling of motivating 
program $\prog$. Figure~\ref{fig:3a} displays the simplified control flow graph 
on which we annotate the memory access related variables. That is, \texttt{S[0]} 
indicates the first array read in the $\mathit{while}$ loop, and \texttt{v1} 
means the memory \emph{write} in the $\mathit{if}$ branch. The solid arrow 
line denotes the execution flow of state $s$, which takes the $\mathit{else}$ 
branch in regular symbolic execution.

In Figure~\ref{fig:3a}, on reaching the branch point, \SpecuSym duplicates a 
\textit{speculative} state $s'$ from $s$. Then, it enforces a bounded symbolic 
execution of $s'$, e.g., one memory access inside the \textit{if} branch, as 
shown by the dashed arrow curve. Once meeting the speculation threshold, 
\SpecuSym stops the exploration and turns back to find state $s$ who is awaiting 
the finish of $s'$. Also, before resuming $s$, \SpecuSym incorporates the cache 
state of $s'$ (cf. Figure~\ref{fig:3b}) into $s$ and terminates $s'$. Right now, 
the entire cache has been filled since the memory write in $s'$ mapped \texttt{v1} 
into line \#256.

Subsequently, executing the memory \emph{store} to \texttt{v2} in state $s$ has
to evict \texttt{S[0]} from cache line \#1 following the LRU policy, as shown in 
Figure~\ref{fig:3b}. Then, the last memory \emph{load} of \texttt{S[x]} might 
result in a cache miss or a hit, counting on the value of \texttt{x}, which is 
consistent with the situation explained in Section~\ref{sec:specu_leak}. Hence, 
\SpecuSym succeeds in modeling speculative execution and conforms to the 
three-phase analysis designed in Section~\ref{sec:specu}. Note that Figure
\ref{fig:3a} only shows the modeling of the \emph{if} branch while \SpecuSym 
considers both branches and misses no potential cases.

\ignore{
In Figure~\ref{fig:3a}, the speculative state $s'$ inherits the cache 
state from $s$, maps \texttt{v2} into a cache line, and merges back the updated 
cache state to state $s$ before its termination, as formally stated in Algorithm
\ref{alg:specusym} (lines 23-25). 
}

In this example, updating the cache state is straightforward. By checking the 
in-cache addresses to find a proper cache line upon a cache miss, or update the 
most recently used line in case of a hit. For the non-deterministic address, i.e., 
\texttt{S[x]}, we can, at worst, try 254 possibilities to test the differences. 
However, eagerly enumerating all potentials is impractical for real-world programs 
due to the unbearable overhead. Furthermore, a great many cache mappings are
redundant in terms of leak exposure. For instance, only 1 specific cache layout 
can reveal the leak in the motivating example.

Instead, \SpecuSym models branch mispredictions along with a path exploration 
and maintains a trace of memory events as the alternative of the cache state. 
Then it analyzes the event trace to build the leak constraint and lazily 
searches for feasible solutions. We detail this approach in Section
\ref{sec:modeling} and Section~\ref{sec:analysis}.

As discussed before, the processor stops speculative execution once it has computed
the branch destination. Since speculation may end at an arbitrary point, we model 
the speculative window with a configurable three-dimensional threshold. That is,
if the number of interpreted events in a \textit{speculative} state reaches the
ROB size or the speculatively executed branch events meet the specified bound, 
we stop speculation immediately. And, if a memory event in speculative execution 
causes a cache miss, we also stop the speculation after its execution. Otherwise, 
we continue the execution.

\subsection{Cache State Modeling}
\label{sec:modeling}


This section addresses how \SpecuSym models the cache state during dynamic
symbolic execution. We extend the definition of a symbolic state $s$ (cf.
Section~\ref{sec:se}) to form a new tuple $\langle \pcon, \mathit{e},
\mathit{brs},\Omega,\pi\rangle$. The newly introduced symbol $\pi$ denotes 
the memory event trace in state $s$. Then we establish the following notions:
\begin{itemize}
  \item A program state in \SpecuSym is either a normal \textit{symbolic} state 
	or a \textit{speculative} state that from the former can the latter be 
	duplicated, but not vice versa. 
  \item A memory event trace denoted as $\pi:=\mathit{\{m_0,...,m_n\}}$, 
    consists of happened memory events in the execution order.
  \item Each memory event $\mathit{m_i}$ in $\pi$, where index $i\in [0,n]$, 
	comes from either a \textit{symbolic} state or a \textit{speculative} state.
  \item $\pi$ and the correlated computations alternatively represent the cache 
	state in a program state $s$.
\end{itemize}
\ignore{
On modeling the cache state, exsiting tools such as CaSym~\cite{BrotzmanLZTK2018} 
maintains and updates cache mappings during symbolic execution; others like 
Chalice~\cite{ChattopadhyayBRZ17} and \textsc{SymSC}~\cite{GuoWW18} uses 
offline constraint solving to to reason about the cache behaviors of memory 
accesses upon concrete cache models. To better coordinate the speculative 
modeling component, \SpecuSym adopts an on-the-fly analysis.
}

To be specific, before interpreting a memory event $m_i$ in a \textit{symbolic} 
state $s$, \SpecuSym analyzes if $\mathit{m_i}$ can cause timing leak under two 
scenarios, namely the new \textit{divergent} behavior and the new \textit{opposite} 
behavior. Note that \SpecuSym won't perform the analyses in any \textit{speculative} 
state. To explain these scenarios, we define four necessary notions as follows.
\begin{itemize}
  \item $\mathit{a_i}$ denotes the used memory address in event $\mathit{m_i}$.
  \item $\mathit{cs(a_i)}$ denotes the cache set that address $\mathit{a_i}$ maps to.
  \item $\mathit{tag(a_i)}$ denotes the unique $\mathit{tag}$ of address $\mathit{a_i}$.
  \item {N} denotes the cache associativity where N$\in$[1, the total number of cache lines] (e.g., N-way set-associative cache). 
\end{itemize}
\subsubsection{New Opposite Behavior}
\label{sec:new_opp}
It means that executing $\mathit{m_i}$ under non-speculative execution causes an 
always-miss, while the result under speculative execution is an always-hit, 
and vice versa. Recall that we only target new leaks introduced by speculative 
execution, and this scene depicts the exact opposite situation. To build the 
formal definition, we define three new notions: 
\begin{itemize}
  \item $\mathit{\eta_{in}(m_i)}$ denotes the condition that $\mathit{m_i}$ 
	triggers an always-hit under speculative execution with input $\mathit{in}$.
  \item $\mathit{\eta^\prime_{in}(m_i)}$ denotes the condition that $\mathit{m_i}$ 
	triggers an always-hit under non-speculative execution with input $\mathit{in}$.
  \item Boolean variable $\mathit{org(m_i)}$ denotes the origin of $\mathit{m_i}$, 
  where 0 means $\mathit{m_i}$ is from a normal \textit{symbolic} state, and 1 
	means $\mathit{m_i}$ comes from a \textit{speculative} state.
\end{itemize}
Then we can formalize $\mathit{\eta_{in}(m_i)}$ and $\mathit{\eta^\prime_{in}(m_i)}$ as follows. 
\begin{multline}
  \label{eqn:hc1}
    \mathit{\eta_{in}(m_i)}~:=~~ 
     \underbrace{ 
      \exists\mathit{j}\in[\mathit{0,i})|
      \wedge 
      \mathit{a_j}=\mathit{a_i} 
    ~\wedge~
      {\forall}x\in\left(j,i\right)\big|\mathit{a_x}\ne\mathit{a_i}
    }_{{\circled{1}}\mathit{~Find~the~nearest~identical~address~a_j}}
		~~~\wedge 
		\\
    ~~~~\sum_{{\scriptscriptstyle \mathit{y=j}+1}}^{{\scriptscriptstyle \mathit{i-}1}} 
    \underbrace{
      \mathit{cs(a_j)}=\mathit{cs(a_y)} 
      \wedge
      \nexists\mathit{z}\in(\mathit{j,y})|\mathit{tag(a_z)}=\mathit{tag(a_y)}
    }_{{\circled{2}}\mathit{~Count~the~unique~a_y~who~and~a_i~map~to~the~same~set}}
    <N
\end{multline}
\vspace{.2ex}
\begin{multline}
  \label{eqn:hc2}
    \mathit{\eta^\prime_{in}(m_i)}:=\mathit{\eta_{in}(m_i)}\wedge
    \underbrace{
      \mathit{org(a_j)}\ne1 
      \wedge
      \mathit{org(a_x)\ne}1 
      \wedge
      \mathit{org(a_y)\ne}1 
    }_{{\circled{3}}\mathit{~Filter~events~from~speculative~execution}}
\end{multline}

Given a $\mathit{m_i}$, we first search $\pi$ for an identical address 
$\mathit{a_j}$, which was visited in event $\mathit{m_j}$ happened before 
$\mathit{m_i}$. Also, $\mathit{a_j}$ has to be the nearest candidate that 
no other addresses between $\mathit{a_j}$ and $\mathit{a_i}$ are qualified
(cf. \circled{1}). Second, we count the unique addresses who and $\mathit{m_i}$
map to the same cache set and assure the number of such addresses is less 
than the set associativity N (cf. \circled{2}).
%

Based on $\mathit{\eta_{in}(m_i)}$ we define $\mathit{\eta^\prime_{in}(m_i)}$
who has an extra constraint (cf. \circled{3}) upon $\mathit{\eta_{in}(m_i)}$. 
The new constraint ensures that all involved memory events must be from 
normal \textit{symbolic} states hence filtering the influences of memory 
events from \textit{speculative} states.

\subsubsection{New Divergent Behavior}
\label{sec:new_div}
It means executing $\mathit{m_i}$ under speculative execution with two different
inputs $\mathit{in}$ and $\mathit{in'}$ can cause a miss and a hit, respectively. 
For instance, reading \texttt{S[x]} exposes this symptom (cf. Section~\ref{sec:specu_leak}). 
To analyze the behavior, we build $\mathit{\mu_{in}(m_i)}$ which returns the cache
hit condition of $\mathit{m_i}$ in trace $\pi$:
\begin{multline}
  \label{eqn:hc}
    \mathit{\mu_{in}(m_i)}:=
    \bigvee_{{\scriptscriptstyle 0{\leq}j<i}}
    \Big(
    \underbrace{
      \mathit{tag(a_j)}=\mathit{tag(a_i)} 
			\wedge 
      \mathit{cs(a_j)}=\mathit{cs(a_i)} 
    }_{{\circled{4}}\mathit{~Identify~the~potential~a_j}}
    ~~~~~~\wedge 
		~~\\
    \underbrace{
      {\forall}x\in\left(j,i\right)\big |\mathit{tag(a_x)}\ne\mathit{tag(a_i)}
      \vee 
      \mathit{cs}(a_x)\ne\mathit{cs}(a_i)
    }_{{\circled{5}}\mathit{~Find~the~nearest~a_j}}
		~~~\wedge 
		\\
    \sum_{{\scriptscriptstyle \mathit{y=j}+1}}^{{\scriptscriptstyle \mathit{i-}1}} 
    \underbrace{
      \mathit{cs(a_j)}=\mathit{cs(a_y)} 
      \wedge
      \nexists\mathit{z}\in(\mathit{j,y})|\mathit{tag(a_z)}=\mathit{tag(a_y)}
    }_{{\circled{6}}\mathit{~Count~the~unique~a_y~who~and~a_i~map~to~the~same~set}}
    <N 
    \Big)
\end{multline}
\vspace{.8ex}

Still, given a $\mathit{m_i}$, we first seek a prior $\mathit{m_j}$ 
from $\pi$ who might let $\mathit{m_i}$ cause a cache hit. That is, the 
addresses $\mathit{a_j}$ and $\mathit{a_i}$ have the same \emph{tag} 
and \emph{set} values (cf. \circled{4}). Moreover, we need to find the 
nearest $\mathit{m_j}$ since $\mathit{m_i}$'s cache behavior directly 
relies on it --- this demand is achieved by ensuring the non-existence 
of a $\mathit{m_x}$ between $\mathit{m_j}$ and $\mathit{m_i}$. And, 
$\mathit{m_x}$ and $\mathit{m_i}$ own the same \textit{tag} and 
\textit{set} values (cf. \circled{5}).

Next, we consider the cache replacement policy. \SpecuSym models the N-way 
set-associative cache with the LRU policy, whereas other policies can be 
modeled and embedded in equation~(\ref{eqn:hc}) as well. Intuitively, on 
finding a candidate $\mathit{m_j}$ for $\mathit{m_i}$, the executed events 
between $\mathit{m_j}$ and $\mathit{m_i}$ should not evict $\mathit{m_j}$ 
from the cache, to promise a cache hit for $\mathit{m_i}$. Under the LRU 
policy, we observe that if satisfying property~\ref{pro:p1}, can this 
non-evict requirement be met.
\begin{pro}
  \label{pro:p1}
  To avoid evicting the most recently used cache line from a cache set 
  \textsc{Set}, the total number of subsequent unique cache mappings to 
  \textsc{Set} must be less than the set associativity.
\end{pro}
Sub-formula \circled{6} checks whether an event $\mathit{m_y}$ between 
$\mathit{m_j}$ and $\mathit{m_i}$ can form a uniquely new mapping to the 
set that $\mathit{a_i}$ maps to. We perform the check by first comparing 
the \textit{set} values of $\mathit{a_y}$ and $\mathit{a_i}$. If the 
comparison satisfies, we further analyze if address $\mathit{a_y}$ has 
never been accessed by an event $\mathit{m_z}$ --- by confirming that the 
\emph{tag} values of $\mathit{a_z}$ and $\mathit{a_y}$ always differ. If 
this check also satisfies, then $\mathit{a_y}$ must form a uniquely new 
cache mapping. We count the new mappings and model \textsc{Property}
\ref{pro:p1} in equation (\ref{eqn:hc}).

Take a 9-event trace $\pi$:=
\{$\mathit{m_1,m_2,\underline{m_1,m_3,m_3,m_4,m_5,m_4,m_1}}$\} as the 
example. For brevity, we assume all the memory addresses associate with 
the same cache set, and each address corresponds to one cache line. And, 
the cache set associativity is \textit{4}. The question is, whether the 
last event, $\mathit{m_1}$, can lead to a cache hit. Here we use 
\#$n$ to index these events where $n\in$[1,9].

Backtracking from the last event $\mathit{m_1}$, we first locate two prior 
$\mathit{m_1}$ events, \#1 and \#3, as the candidates that satisfy \circled{4}. 
Then, according to \circled{5}, we select the nearest event \#3, to form a 
sub-trace $\rho$ as underlined in $\pi$. Note that after $\rho$'s head event 
$\mathit{m_1}$, the cache line used by this $\mathit{m_1}$ becomes the most 
recently used line. Let's name the line as \emph{l}. Then, following \circled{6}, 
along with $\rho$ we can identify \emph{3} unique new cache mappings incurred 
by events \#4 ($\mathit{m_3}$), \#6 ($\mathit{m_4}$), and \#7 ($\mathit{m_5}$), 
respectively. Other events, i.e., \#5 ($\mathit{m_3}$) and \#8 ($\mathit{m_4}$),
cannot form unique mappings since they do not satisfy \circled{6}.

Therefore, we conclude that the target event, $\mathit{m_1}$, must lead to a 
cache hit because the identified \emph{3} unique cache mappings at most evict 
the other 3 lines rather than the line \emph{l}. In other words, the number of 
uniquely new mappings is less than the cache set associativity, which conforms 
to \textsc{Property}~\ref{pro:p1}.

\subsection{Cache Behavior Analysis}
\label{sec:analysis}

This section leverages the formal terms $\mathit{\mu_{in}(m_i)}$, 
$\mathit{\eta_{in}(m_i)}$ and $\mathit{\eta^\prime_{in}(m_i)}$ (cf. Section
\ref{sec:modeling}) to analyze the cache behavior of $\mathit{m_i}$. In general, 
if $\mathit{m_i}$ has no cache behavior variance in both speculative execution 
and non-speculative execution for any inputs, it has no leaks. Otherwise, at the 
program location of $\mathit{m_i}$, there is a leak.


In Algorithm~\ref{alg:specusym}, \SpecuSym invokes $\mathit{AnalyzeCache}$ right
before interpreting a memory event (line 13). In this procedure, \SpecuSym first 
examines whether the current state $s$ is a $\mathit{symbolic}$ state, and the 
event $s.e$ relates to sensitive input (line 30). It is because the timing effect 
inside speculative execution is externally invisible, and we are interested in the 
sensitive data relevant memory accesses. Next, \SpecuSym builds the leak constraint 
$\theta$ for event $s.e$ and solves for a solution (lines 31-33). If the solution
does not exist, \SpecuSym claims leakage-free at $s.e$ in current state $s$.

To form $\theta$, we need to build the constraint exposing new cache behaviors
at a given memory event $\mathit{m_i}$ as follows.
\begin{multline}
  \label{eqn:new_opp}
  ~~~~~~~
  \mathit{opp_i}:=~
  \forall\mathit{in,in'}.
  \big(
  \mathit{\eta_{in}(m_i)\neq\eta_{in'}^\prime(m_i)}
  \big)
  ~~~~~~~~~~~~~~~~~~~~~~~
\end{multline}
\begin{multline}
  \label{eqn:new_div}
  ~~~~~~~
  \mathit{div_i}:=~
  \exists\mathit{in,in'}.
  \big(
  \mathit{in\neq in'\wedge\mu_{in}(m_i) \neq \mu_{in'}(m_i)}
  \big)
  ~~~~~~~
\end{multline}
\ignore{
\begin{multline}
  \label{eqn:leak}
  ~~
  \theta:=
  \Big(
  \mathit{div_i}
  \wedge
  \big(
  \sum_{\scriptscriptstyle \mathit{k=0}}^{\scriptscriptstyle\mathit{n}} 
  \mathit{\eta_{in''}(m_k)=}0
  \neq
  \sum_{\scriptscriptstyle \mathit{k=0}}^{\scriptscriptstyle\mathit{n}} 
  \mathit{\eta_{in''}(m_k)=}1
  \big)
  \Big)
  \vee 
  \mathit{opp_i}
  ~~
\end{multline}
}
\ignore{
Note that 
speculative execution not only evicts cache lines to cause new misses but also loads 
memory data into cache that benefits new hits. Since a leak refers to the noticeable 
timing variance in program runs with and without speculative execution, from 
$\mathit{opp_i}$ the amount of new misses should differ from the amount of new hits. 
And we construct the final leak constraint $\theta$ on this observation:
}
Constraint $\mathit{opp_i}$ checks the existence of \textit{New Opposite Behavior}
and $\mathit{div_i}$ reasons \textit{New Divergent Behavior}, respectively. We 
define $\theta:=\mathit{opp_i}\vee\mathit{div_i}$ as the leak constraint at 
$\mathit{m_i}$. \SpecuSym first computes $\mathit{\eta_{in'}^\prime(m_i)}$ 
because of the lower cost of a \textit{must-be} solving. Then, it uses the 
concretized value to substitute $\mathit{\eta_{in'}^\prime(m_i)}$ in 
$\mathit{opp_i}$ for the further solving of $\mathit{in}$. If this targeted 
$\mathit{in}$ exists, then $opp_i$ must be satisfiable, and \SpecuSym can 
safely skip the rest analysis of $\mathit{div_i}$. Otherwise, it has to 
continue reasoning $\mathit{div_i}$. Finally, if the solver successfully 
returns a concrete solution of $\theta$, \SpecuSym generates a test case 
including two inputs, the trace $\pi$, and event $\mathit{m_i}$ as the leak 
witness.

For instance, in Figure~\ref{fig:3a}, before interpreting the \texttt{load} 
event of \texttt{S[x]}, \SpecuSym analyzes the cache since the sensitive 
data relevant address {\texttt{S[x]}} is from a \textit{symbolic} state 
$s$. At this point, trace $\pi$ in $s$ is {\small \{\texttt{S[0]},...,
\texttt{S[254]},\texttt{x},\texttt{v2},\texttt{v1},\texttt{S[x]}\}}. Also,
in non-speculative execution reading {\small\texttt{S[x]}} must get a cache 
hit (cf. Section~\ref{sec:no_leak}). 
Substituting $\mathit{\eta_{in'}^\prime(m_i)}$ with value 1 in $\mathit{opp_i}$, 
\SpecuSym tries to solve $\forall\mathit{in}.(\mathit{\eta_{in}(m_i)\neq}$1). 
Unfortunately, this simplified $\mathit{opp_i}$ is still unsatisfiable since 
$\mathit{\eta_{in}(m_i)}$ evaluates to 1 if $\mathit{in}$$\neq$0. Therefore, 
\SpecuSym moves forward to $\mathit{div_i}$. This time, it successfully returns 
two values 0 and 1, which expose a cache miss and a hit, respectively. Finally, 
\SpecuSym outputs the solved values, the trace $s.\pi$, and the \textit{read} 
event of {\texttt{S[x]}}.

It is worth noting that processors may mispredict branches several times 
along a program path. As a result, we need to count the cache side effects 
from multiple mispredictions. However, attackers always expect the minimum 
effort to trigger mispredictions for leak purposes. Meanwhile, manipulating 
multiple speculation windows is very difficult for external attackers. In 
\textsc{SpecuSym}, we follow the attacker's perspective to study whether one 
misprediction could be enough to cause timing leaks and leave the multiple 
speculation cases as our future work.

\subsection{Optimizations}
\label{sec:optimize}

\subsubsection{Constraint Transformation.}
In Section~\ref{sec:analysis}, constraints $\mathit{opp_i}$ and $\mathit{div_i}$
capture the complete set of leak-related cache behaviors. However, based on the
timing-leakage-free assumption of a target program under non-speculative execution,
we observe a property which helps \SpecuSym accelerate the cache analysis:
\begin{pro}
  \label{pro:p2}
  If under the same input $\mathit{in:=\{\lambda,t\}}$, the cache behaviors of a 
	$\lambda$-related	memory event $\mathit{m_i}$ in speculative execution and 
	non-speculative execution are opposite, there is a cache timing leak. 
\end{pro}
Property~\ref{pro:p2} reflects the intersection of $\mathit{opp_i}$ and 
$\mathit{div_i}$. As assumed, the cache behavior of $\mathit{m_i}$ in 
non-speculative execution is deterministic. In case an opposite behavior 
under speculative execution, $\mathit{m_i}$ would cause either new opposite 
behavior or new divergent behavior. In any case, it is a leak. Then the 
solving of $\theta$ turns to be reasoning the existence of such $\mathit{in}$, 
which has cheaper computation cost. We model the property into a new constraint 
$\theta'$:  
\begin{multline}
  \label{eqn:leak}
  ~~~~~~~~~~~~~~~
  \theta':=
  \exists\mathit{in}.
  \mathit{
    \mu_{in}(m_i)~\neq~\eta_{in}^\prime(m_i)
  }
  ~~~~~~~~~~~~~~~
\end{multline}
Due to the assumption mentioned above, $\mathit{\eta_{in}^\prime(m_i)}$ 
is always 1 (hit) or 0 (miss). Accordingly, if $\mathit{\mu_{in}(m_i)}$
(may-hit condition under speculation) evaluates to 0 or 1 on the same 
$\mathit{in}$, either $\mathit{opp_i}$ or $\mathit{div_i}$ may evaluate 
to 1, which already deduces a leak without solving $\mathit{opp_i}$ or 
$\mathit{div_i}$. Therefore, if $\theta'$ is satisfiable, $\theta$ also 
satisfies.

\subsubsection{Formula Reduction.}

However, the size of $\theta'$ still intensely increases along with the 
increment of $i$, which pressures the constraint solver. By inspecting 
the constraint structure, we separate lengthy formulas into smaller 
pieces by following strategies. First, if a subformula of a Conjunctive 
Normal Form (CNF) formula evaluates as \textit{false}, we directly return 
\textit{false}. For example, in equation (\ref{eqn:hc}), if \circled{4} 
is \textit{false}, we skip querying solver for \circled{5} and \circled{6}. 
Second, if a subformula in a Disjunction Normal Form (DNF) formula is 
determined as \textit{true}, we discard all other subformulas. As in 
equation (\ref{eqn:hc}), once we find the nearest $\mathit{a_j}$ by 
\circled{5}, we discard subformulas for indices before $j$. Third, if 
two subformulas are in negated forms, we avoid an extra solver query. For 
example, in equation~\ref{eqn:hc}, constraint \circled{5} of $\mathit{m_j}$ 
is the negation of constraint \circled{4} of $\mathit{m}_{\mathit{j}+1}$, 
which implies we can save one constraint solving time. Fourth, since we 
repeatedly traverse memory event trace in constraint construction, caching 
the intermediate results avoids repeated computation. We bank the computed 
formulas into a hash map so that \SpecuSym can quickly retrieve them with 
tolerable storage overhead.

\subsubsection{Speculative Assumption Checking.}
Recall that in Algorithm~\ref{alg:specusym}, \SpecuSym duplicates a new state 
$s'$ from symbolic state $s$ (line 21) if the branch predicate closely relates 
to a memory value rather than a cache value. As mentioned earlier in Section
\ref{sec:modeling}, this strategy filters transient speculations that are unlikely 
to affect the cache. Wu et al.~\cite{WuW19} over-approximately treated all memory 
events as actual memory visits, though predicate-related values may have been 
loaded to the cache. \SpecuSym proposes an alternative for symbolic execution to 
deal with this issue. Ideally, to address the problem, we have to perform a 
burdensome backward cache analysis from the current branch to the first memory 
access, examining whether the value involved in the branch predicate is in cache 
or not. However, we observe that the memory accesses from the current branch 
to the last branch typically have a critical influence on the cache state before 
the current branch. Therefore, to balance the precision and efficiency, \SpecuSym 
backwardly examines the memory accesses in this range. If the predicate-related 
values have been fetched to cache beforehand and still exist in the cache, we 
regard the speculative modeling at this branch as unnecessary. Otherwise, we still 
conduct all analyses as usual, as usual, to avoid missing leaks. Although this 
optimization introduces extra computation overhead, it still benefits overall 
performance since the targeted cases commonly exist in programs because of the time 
and spatial locality purposes.

\section{Evaluations}
\label{sec:evaluation}

We have implemented \SpecuSym based on the KLEE~\cite{CadarDE08} symbolic 
executor and the LLVM compiler~\cite{LattnerA04}. We refit KLEE in three
main aspects. First, we made KLEE support the bounded execution of auxiliary 
\textit{speculative} state and schedule the parent \textit{symbolic} state 
afterward, to mimic the branch misprediction. Second, we made KLEE stitch 
a sequence of memory events from both normal \textit{symbolic} state and 
\textit{speculative} states, to represent the incorporation of cache data 
from speculation. Third, we made KLEE analyze the cache behaviors on memory 
events to generate concrete test cases for timing leaks. Based on these new 
components, we built \SpecuSym for the cache timing leak analysis under 
speculative execution.

Specifically, after loading the LLVM bit-code of the target program, \SpecuSym 
executes it symbolically, explores speculative states, records the memory 
event trace, and conducts cache analysis following Algorithm~\ref{alg:specusym}. 
The leak constraint is encoded in Z3-compatible form. On solving the constructed 
constraints, \SpecuSym outputs leak witness, including the event trace, the leaky 
memory operation location, and the solved inputs.

We design the following research questions for the experiments: 
\begin{itemize}
\item Can \SpecuSym successfully identify cache timing leaks introduced 
by speculative execution?
\item Can \SpecuSym complement the abstract interpretation based method 
by providing more accurate results?
\item Can the proposed optimizations boost the overall performance of \textsc{SpecuSym}?
\end{itemize}

\subsection{Benchmark Programs}
\label{sec:benchs}

\begin{table}
\caption{\fontsize{8}{11}\selectfont Benchmark statistics: Name, Lines 
of C code (LoC), Source, Sensitive Input Size in Bytes (S-In), and Number of 
Branches (Brs).}
\label{tbl:benchs}
\centering
\resizebox{\linewidth}{!}{
\begin{tabular}{|rrlcr|} 
\hline
\multicolumn{1}{|c}{\textbf{Name}} & \multicolumn{1}{c}{\textbf{LoC}} &\multicolumn{1}{c}{\textbf{Source}}&\multicolumn{1}{c}{\textbf{S-In}}&\multicolumn{1}{c|}{\textbf{Brs}}\\
\hline
\texttt{hash}\cite{hpn-ssh} & 320 & The hpn-ssh hash implementation & 64 & 5 \\
\hline
\texttt{AES}\cite{LibTomCrypt} & 1,838 & The LibTomcrypt AES cipher & 16 & 17 \\
\texttt{blowfish}\cite{LibTomCrypt} & 467 & The LibTomcrypt blowfish cipher& 8& 11 \\
\texttt{chacha20}\cite{LibTomCrypt} & 776 & The LibTomcrypt chacha20 cipher& 36& 133 \\
\texttt{encoder}\cite{LibTomCrypt} & 134 & The LibTomcrypt hex encoder& 100&2  \\
\texttt{ocb}\cite{LibTomCrypt} & 377 & The LibTomcrypt OCB implementation& 28&23 \\
\hline
\texttt{DES}\cite{OpenSSL111c} & 1,100 & The OpenSSL DES cipher& 64&79 \\
\texttt{str2key}\cite{OpenSSL111c} & 371 & The OpenSSL key prepare for DES& 16&8 \\
\hline
\texttt{DES}\cite{glibc} & 547 &The glibc DES implementation & 16 &7 \\
\hline
\texttt{Camellia}\cite{Tegra}& 1,324  & The NVIDIA Tegra Camellia cipher&16&16 \\
\texttt{Salsa}\cite{Tegra}& 279  & The NVIDIA Tegra Salsa20 stream cipher&28&12 \\
\texttt{Seed}\cite{Tegra}& 487  & The NVIDIA Tegra Seed cipher&16&2 \\
\hline
\texttt{DES}\cite{Libgcrypt} & 337&The Libgcrypt DES cipher&8&3 \\
\texttt{Salsa}\cite{Libgcrypt} &344&The Libgcrypt Salsa20 stream cipher&40&22 \\ 
\hline
\texttt{Spectre}\cite{spectre-attack} &90&The Spectre V1 application& 4&1 \\ 
\hline
\end{tabular}
}
\end{table}

Table~\ref{tbl:benchs} shows the statistics of the benchmarks for evaluation.  
The first three columns, \textbf{Name}, \textbf{LoC}, and \textbf{Source}, 
indicate the names, the lines of code, and the sources of these benchmarks. 
Column \textbf{S-In} denotes the sensitive input size in bytes. The last 
column \textbf{Brs} shows the number of conditional branches in each program.

Our benchmark suite consists of a diverse set of open-source C programs. 
The first program comes from hpn-ssh~\cite{hpn-ssh}. The second group has 
five programs from LibTomCrypt 1.18.1~\cite{LibTomCrypt}. The third group 
uses two programs from glibc 2.29~\cite{glibc}. The fourth group includes 
three programs from the Tegra library~\cite{Tegra}. The rest benchmarks 
are from the Libgcrypt 1.8.4~\cite{Libgcrypt} and the Spectre vulnerability
application~\cite{spectre-attack}. Most benchmarks are computation-intensive 
despite the compact program volumes. The sensitive input of each benchmark 
is initialized to a symbol whose size shows in column \textbf{S-In}. Also, 
each program has 1 to 133 conditional branches.

We evaluate the benchmarks on the N-way set-associative cache and LRU policy 
with different cache settings. To be specific, we design four caches as (1) 
32KB cache size with N=4; (2) 32KB cache size with N=8; (3) 64KB cache size 
with N=8; and (4) 32KB cache size with N=512. Each cache line has 64 bytes 
for all caches. The first three settings are close to the L1 data cache 
classes in modern processors. The last setting creates a fully-associative 
cache on which we compare \SpecuSym with \cite{WuW19}. Besides, we evaluate 
the effectiveness of optimizations using setting (3). All the evaluations 
are conducted on a machine running Ubuntu 16.04 64-bit Server Linux with 
Intel(R) Xeon(R) 2.20GHz CPUs 24 cores and 256GB RAM. Each benchmark is 
allowed to run at most 12 hours.

\subsection{Experimental Results}
\label{sec:experiments}

\subsubsection{Timing Leak Detection}
\label{sec:leak_detect}

\begin{table}
\caption{Detection results on set-associative caches}
\label{tbl:detection}
\centering
\resizebox{\linewidth}{!}{
\begin{tabular}{rcccccc}
\toprule
\multirow{2}{*}{\textbf{Name~~~~}} 		&  
\multicolumn{2}{c}{\textbf{$C_1$: 32KB, 4-Way}}&		
\multicolumn{2}{c}{\textbf{$C_2$: 32KB, 8-Way}}&		
\multicolumn{2}{c}{\textbf{$C_3$: 64KB, 8-Way}}\\
\cmidrule(r){2-3} \cmidrule(r){4-5} \cmidrule(r){6-7}
&Time (m)  &\#.D/O &Time (m)  &\#.D/O &Time (m) & \#.D/O\\ 
\hline
\texttt{hash}\cite{hpn-ssh} & 0.57 & 0/0 & 0.56 & 0/0 & 0.56 & 0/0 \\
\hline
\texttt{AES}\cite{LibTomCrypt} &  32.53 & 20/3 & 99.43 & 25/0 & 65.27 & 26/0 \\ 
\texttt{blowfish}\cite{LibTomCrypt} & 0.04 & 0/0 & 0.03 & 0/2 & 0.03 & 0/0 \\ 
\texttt{chacha20}\cite{LibTomCrypt} & 0.31 & 0/5 & 0.31 & 0/61 & 0.28 & 0/61 \\ 
\texttt{encoder}\cite{LibTomCrypt} & < 0.01 & 0/5 & < 0.01 & 0/5 & < 0.01 & 0/5 \\ 
\texttt{ocb}\cite{LibTomCrypt} & < 0.01 & 0/0 & < 0.01 & 0/0 & < 0.01 & 0/0 \\ 
\hline
\texttt{DES}\cite{OpenSSL111c} & 10.72 & 0/5 & 10.65 & 0/10 & 10.77 & 0/10 \\ 
\texttt{str2key}\cite{OpenSSL111c} & 0.01 & 0/0 & < 0.01 & 0/0 & < 0.01 & 0/0 \\ 
\hline
\texttt{DES}\cite{glibc} & 51.98  & 16/6 & 384.57 & 26/1 & 393.48 & 24/3 \\ 
\hline
\texttt{Camellia}\cite{Tegra} & 0.13 & 0/0 & 0.12 & 0/0 & 0.13 & 0/0 \\ 
\texttt{Salsa}\cite{Tegra} & 0.03 & 0/0 & 0.03 & 0/0 & 0.04 & 0/0 \\ 
\texttt{Seed}\cite{Tegra} & 0.21 & 0/0 & 0.21 & 0/0 & 0.21 & 0/0 \\ 
\hline
\texttt{DES}\cite{Libgcrypt} & 0.37 & 0/0 & 0.37 & 0/0 & 0.37 & 0/0 \\ 
\texttt{Salsa}\cite{Libgcrypt} & < 0.01 & 0/0 & < 0.01 & 0/0 & < 0.01 & 0/0 \\ 
\hline
\texttt{Spectre}\cite{spectre-attack}& <0.01 & 0/0 & < 0.01 & 0/0 & < 0.01 & 0/0 \\ 
\bottomrule
\end{tabular}
}
\vspace{-2ex}
\end{table}

In this section, we conduct experiments for the first research question. Table
\ref{tbl:detection} shows the leak detection results under the 3 set-associative 
cache settings. Let's name these caches as {$C_1$}, $C_2$, and $C_3$. $C_1$ has 
a 32KB size, and each cache set consists of 4 lines. As each line has 64 bytes, 
$C_1$ has 128 sets in total. $C_2$ also owns 32KB size, but its associativity 
increases to 8. Hence it has 64 sets. $C_3$ has 64KB size while its associativity 
remains 8, thereby it has 128 sets and 1024 cache lines. For each benchmark, we 
collect execution statistics under all 3 cache settings, including the total 
execution time in minute (Time (m)) and the amount of divergent/opposite leaks,
named \#.D/O.

Overall, \SpecuSym identified that 6 out of 15 benchmarks suffer from timing 
leaks. Specifically, \texttt{blowfish} is leaky only under cache $C_2$ while 
the other 5 programs have leaks under all 3 caches. Among these 5 programs, 
\texttt{DES}~\cite{glibc} has both D/O leaks in all caches; \texttt{AES} has 
both D/O leaks in $C_1$, but only opposite leaks in $C_2$ and $C_3$; and the 
rest 3 programs (\texttt{chacha20}, \texttt{encoder}, and \texttt{DES}
\cite{OpenSSL111c}) have only opposite leaks under all the 3 caches.

Next, we compare the results under $C_1$ and $C_2$ to research how the cache 
associativity variation affects leak occurrence. Intuitively, the more lines 
per cache set, the less potential cache conflicts hence fewer leaks. However, 
experiments affirm that the increased associativity indeed triggers more leaks. 
First, \SpecuSym detected 60 and 130 leaks under $C_1$ and $C_2$, respectively; 
second, 5 out of the 6 leaky programs have more leaks under $C_2$. The only 
exception, \texttt{encoder}, has 5 leaks in both $C_1$ and $C_2$. Third, the 
non-leaky program under $C_1$, \texttt{blowfish}, now turns to expose 2 new 
leaks under $C_2$. This phenomenon is mostly because the speculative memory 
visits raise more cache hits --- the misses under non-speculative execution 
now have higher possibilities to be cache hits since speculative execution has
fetched memory data earlier into the cache.

Further, we compare between $C_2$ and $C_3$, to study how the cache size could 
affect leak detection. We find that despite the sizeable 32KB capacity 
difference, the detection results show no drastic variances. First, \SpecuSym 
detected the same types and amounts of leaks in 3 programs (\texttt{chacha20}, 
\texttt{encoder}, and \texttt{DES}~\cite{OpenSSL111c}) in both $C_2$ and $C_3$. 
The analysis time is also close. Second, another two leaky programs, \texttt{AES} 
and \texttt{DES}~\cite{glibc}, also exhibit similar statistics. Third, the main 
difference comes from \texttt{blowfish}, which has two leaks on $C_2$ but no 
leaks on $C_3$. Though this unique case indicates that smaller cache might 
incur more leaks, it is not the majority situation.

Besides, we observe drastically varying results from different versions of the 
same algorithm. For example, in three DES implementations from OpenSSL
\cite{OpenSSL111c}, glibc~\cite{glibc}, and Libgcrpt~\cite{Libgcrypt}, 
\SpecuSym detected only opposite leaks in OpenSSL \texttt{DES}, both D/O 
leaks in glibc \texttt{DES}, and no leaks in Libgcrypt \texttt{DES}. Meanwhile, 
OpenSSL \texttt{DES} needs about 11 minutes of analysis time upon all caches. 
The glibc \texttt{DES} uses 52 minutes for $C_1$ but around 6.5 hours in both 
$C_2$ and $C_3$. By contrast, Libgcrypt \texttt{DES} uses merely 1 minute upon 
all 3 caches.

Therefore, we can answer the first research question: \SpecuSym is capable of 
detecting timing leaks introduced by speculative execution. It also supports 
various caches for comparative analysis.

\subsubsection{Existing Method Comparison}
\label{sec:comparison}

We compare \SpecuSym with \cite{WuW19} in Table~\ref{tbl:cmp}, to answer the 
second research question. Column 1 of Table~\ref{tbl:cmp} lists all the used
benchmarks in~\cite{WuW19} for timing leak evaluation. Columns 2-3 show the 
detection result and the computation time of this abstract interpretation 
based method~\cite{WuW19}. Columns 4-6 depict the result of \textsc{SpecuSym}, 
while column 5 lists the total amount of detected leaks. To ensure a fair 
comparison, we configure a 512-line fully-associative cache with LRU policy, 
which equals the cache setting used in~\cite{WuW19}. Let's name it as $C_4$.

\begin{table}
\caption{Comparison betwee Wu et al.~\cite{WuW19} and \SpecuSym}
\label{tbl:cmp}
\centering
\resizebox{\linewidth}{!}{
\begin{tabular}{rccccc}
\toprule
\multirow{2}{*}{\textbf{Name~~~~}} 		&  
\multicolumn{2}{c}{\textbf{Wu et al.}~\cite{WuW19}}&		
\multicolumn{3}{c}{\textbf{SpecuSym}}\\
\cmidrule(r){2-3} \cmidrule(r){4-6} 
&Leak Detected   &Time (s) &Leak Detected & \#.Leaks  &Time (s) \\ 
\hline
\texttt{hash}\cite{hpn-ssh} &\underline{~~~\ding{52}~~~} &1.15  & \underline{~~~\ding{55}~~~}  &0  &33.61\\ 
\texttt{AES}\cite{LibTomCrypt} &\ding{55} &2.13 &\ding{55}  &0  &5.01\\ 
\texttt{chacha20}\cite{LibTomCrypt} &\ding{52} &9.24 & \ding{52} &9  &17.84\\ 
\texttt{encoder}\cite{LibTomCrypt} &\ding{52} &0.10 & \ding{52} &5 &0.13\\ 
\texttt{ocb}\cite{LibTomCrypt} &\underline{~~~\ding{52}~~~} &0.68 & \underline{~~~\ding{55}~~~} &0  & 0.05\\ 
\texttt{DES}\cite{OpenSSL111c} &\ding{52} &14.20 & \ding{52}   &8  &725.66\\ 
\texttt{str2key}\cite{OpenSSL111c} &\ding{55} &0.01 &\ding{55}   &0  &0.69\\ 
\texttt{Camellia}\cite{Tegra} &\ding{55} &6.35 &\ding{55}  &0  &7.59\\ 
\texttt{Salsa}\cite{Tegra} &\ding{55} &0.06 &\ding{55}  &0  &0.99\\ 
\texttt{Seed}\cite{Tegra} &\ding{55} &0.07 &\ding{55}   &0  &12.15\\ 
\bottomrule
\end{tabular}
}
\end{table}

Overall, \cite{WuW19} reported that 5 of the 10 benchmarks are leaky, as marked 
with \ding{52} in column 2. Consequently, the rest 5 programs are free of timing 
leaks. For these robust programs, \SpecuSym also detects no leaks, as shown in 
column 4. The result means \SpecuSym introduces no false positives on analyzing 
these programs. In contrast, for those \ding{52} results from~\cite{WuW19}, 
\SpecuSym provides different results on two benchmarks \texttt{hash} and \texttt{ocb}, 
as underlined in Table~\ref{tbl:cmp}. They are deemed to be leaky in~\cite{WuW19},
while \SpecuSym found no leaks in them. We manually inspect the two programs and 
confirm that \SpecuSym gives the right answer --- no leaks exist in these programs. 
The inaccurate results of~\cite{WuW19} should root from its over-approximation 
solution, and they are indeed false positives.

For the remaining programs, \texttt{chacha20}, \texttt{encoder}, and \texttt{DES}, 
both \SpecuSym and~\cite{WuW19} have detected leaks. However, \cite{WuW19} only 
reports property violations, whereas \SpecuSym generates rich information. For 
example, \SpecuSym pinpoints 9, 5, and 8 leaky sites in these benchmarks, respectively. 
Moreover, \SpecuSym can provide inputs and memory event trace for leak diagnosis. 
Note that \SpecuSym generally performs faster on $C_4$ than on other three 
set-associative cachess since the extreme cache setting of $C_4$ advances both cache 
utilization and cache hit rate for these benchmarks.

Compared to~\cite{WuW19}, \SpecuSym may have false negatives because of two reasons. 
First, bounded speculative modeling might miss certain speculation cases. Second, \SpecuSym 
stops speculation once it confirms that a memory event must cause a cache miss, which 
skips the execution of the following events. However, \SpecuSym won't have false 
positives because of high-fidelity behavior modeling and fine-grained per-event 
reasoning approaches.

Another drawback of \SpecuSym is the computation overhead. Table~\ref{tbl:cmp} shows 
that \cite{WuW19} finishes all benchmarks in 35 seconds, whereas \SpecuSym needs 
725.7 seconds to analyze an OpenSSL {\tt{DES}}. Worse, the analysis time under 
$C_3$ for the glibc \texttt{DES}~\cite{glibc} even rises to about 6.6 hours (cf.
Table~\ref{tbl:detection}). The cost mainly comes from constraint solver since 
\SpecuSym reasons individual memory accesses on each program path. However, 
sacrificing the performance for the precision achieves reasonable paybacks. First, 
the increased overhead of most benchmarks is still tolerable. Next, \SpecuSym finds 
false positives from~\cite{WuW19} and generates precise inputs. Moreover, if
\cite{WuW19} provides assisting information, e.g., the suspicious locations on the 
control flow graph, \SpecuSym can purposely guide its analysis to the problematic 
area at an earlier stage.

Now, we answer the second research question: \SpecuSym can complement the latest 
abstract interpretation based analysis~\cite{WuW19}. It introduces no false negatives 
and identifies two false positives reported in~\cite{WuW19}. Further, it can provide
inputs, the speculation flows, and the leak locations for detailed diagnosis.

\subsubsection{Optimization Performance}
\label{sec:opt_perform}

This section studies the performance of the optimizing strategies proposed in 
Section~\ref{sec:optimize}. Table~\ref{tbl:opt} compares the overall performance 
between the baseline \SpecuSym without any optimizations (\texttt{Base}), and the
optimized \SpecuSym with all optimizations enabled (\texttt{Opt}), as listed in 
column 1. Column 2 and column 3 compare the analysis time and the size of involved 
constraints between \texttt{Base} and \texttt{Opt}. Column 4 and column 5 disclose 
the number of identified new divergent leaks (\#.\textbf{Div}) and new opposite 
leaks (\#.\textbf{Opp}), respectively. The cache setting for this experiment is
$C_3$, under which \texttt{Base} and \texttt{Opt} have the minimum performance 
gap, forming a conservative comparison result.

As shown in column 2, after optimizations, the analysis time of the \texttt{Opt} 
\SpecuSym significantly decreases, falling to only $30.7\%$ of the analysis time
in \texttt{Base}, which confirms the effectiveness of the optimizations in 
accelerating the analyses.

Apart from the time drop, the \texttt{Opt} \SpecuSym reduces the size of the 
involved constraint by more than $20\%$, compared to the \texttt{Base} version, 
as shown in column 3. Though the reduction to constraint sizes looks not as 
impressive as the time reduction, the proposed strategies greatly ease the solver 
burden by constraint transformation and formula reduction. Undoubtedly, constraint 
size reduction is the primary reason for the time decrease.

Moreover, column 4 and column 5 disclose that the \texttt{Opt} \SpecuSym 
detects 31.8\% more leaks in total than the \texttt{Base} version, within 
a 12-hour threshold. In particular, \texttt{Opt} identifies 3 times more 
divergent leaks than the \texttt{Base} version, which highlights the importance
of our proposed optimization strategies.

\tikzset{align at top/.style={baseline=(current bounding box.north)}}
\begin{figure}
\subfigure[The benchmark analysis time]{
	\label{fig:4a}
	\begin{minipage}{0.486\linewidth}
	\centering
		\resizebox{1.0\linewidth}{!}
		{
		\begin{tikzpicture}
		\begin{axis}[ 
			ylabel= Time (min), 
			xtick= {1,2,3,4,5,6,7,8,9,10,11,12,13,14,15},
			xticklabel style={rotate=45, anchor=north east, inner sep=2mm},
			xticklabels = {hash,AES,blowfish,chacha20,encoder,ocb,DES,str2key,DES,Camellia,Salsa,Seed,DES,Salsa,Spectre},
			ytick = {0, 200, 400, 600, 800}, 
			yticklabels = {0, 200, 400, 600, 800}, 
			xticklabel style = {font=\large}, 
			yticklabel style = {font=\Large},
			xlabel style = {font=\Large},
			ylabel style = {font=\Large},
			ymin = 0,
			ymax = 800,
			legend style = {font=\Large}, 
			legend pos=north east, 
			grid=both]

			\addplot [line width=1pt, , color = blue, mark=triangle] plot coordinates {
					(1, 3.56) 
						(2, 720) 
						(3, 2.57) 
						(4, 61.42) 
						(5, 0.01) 
						(6, 0.00) 
						(7, 416.58) 
						(8, 0.73) 
						(9, 720) 
						(10, 48.108) 
						(11, 3.71) 
						(12, 290.68) 
						(13, 0.42) 
						(14, 0.02) 
						(15, 0.00) 
				};
		\label{plot:p1}
		\addplot [line width=1pt, , color = red, mark=*] plot coordinates {
			(1, 0.56)
				(2, 32.21)
				(3, 0.04)
				(4, 0.31)
				(5, 0.00)
				(6, 0.00)
				(7, 10.71)
				(8, 0.01)
				(9, 52)
				(10, 0.13)
				(11, 0.03)
				(12, 0.21)
				(13, 0.37)
				(14, 0.00)
				(15, 0.00)
		};
		\label{plot:p2}
		\addlegendentry{\texttt{Base}}
		\addlegendentry{\texttt{Opt}}
		\end{axis}
		\end{tikzpicture}
	}
	\end{minipage}\hfill%
}
\subfigure[The SMT constraint size]{
	\label{fig:4b}
	\begin{minipage}{0.486\linewidth}
	\centering
		\resizebox{1.0\linewidth}{!}{
			\begin{tikzpicture}
			\begin{axis}[ 
				ylabel= Size (\# of constructs), 
				xtick= {1,2,3,4,5,6,7,8,9,10,11,12,13,14,15},
				xticklabel style={rotate=45, anchor=north east, inner sep=2mm},
				xticklabels = {hash,AES,blowfish,chacha20,encoder,ocb,DES,str2key,DES,Camellia,Salsa,Seed,DES,Salsa,Spectre},
				ytick = {0, 1000, 2000, 3000, 4000}, 
				yticklabels = {0, 1000, 2000, 3000, 4000}, 
				xticklabel style = {font=\large}, 
				yticklabel style = {font=\Large},
				xlabel style = {font=\Large},
				ylabel style = {font=\Large},
				ymin = 0,
				ymax = 4000,
				legend style = {font=\Large}, 
				legend pos=north east, 
				grid=both]
				\addplot[line width=1pt, , color = blue, mark=triangle] plot coordinates {
				(1, 1003)
					(2, 2964)
					(3, 0)
					(4, 8)
					(5, 16)
					(6, 0)
					(7, 17)
					(8, 44)
					(9, 3816)
					(10, 2265)
					(11, 0)
					(12, 2382)
					(13, 457)
					(14, 0)
					(15, 111)
				};
				\label{plot:p3}
				\addplot[line width=1pt, , color = red, mark=*] plot coordinates {
					(1, 949)
					(2, 524)
					(3, 0)
					(4, 8)
					(5, 16)
					(6, 0)
					(7, 17)
					(8, 25)
					(9, 2207)
					(10, 1972)
					(11, 0)
					(12, 321)
					(13, 273)
					(14, 0)
					(15, 111)
				};
				\label{plot:p4}
			\addlegendentry{\texttt{Base}}
			\addlegendentry{\texttt{Opt}}
			\end{axis}
			\end{tikzpicture}
		}
	\end{minipage}\hfill
}
\vspace{-2ex}
\caption{Performance comparison for the individual program. Benchmarks are listed by their orders in Table~\ref{tbl:benchs}.}
\label{fig:InsampleOutsample}
\label{fig:opt}
\end{figure}
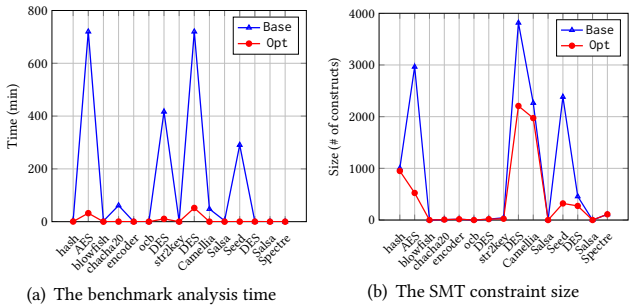

\begin{table}
\caption{The performance boost due to optimizations. }
\vspace{-0.5em}
\label{tbl:opt}
\centering
\resizebox{\linewidth}{!}{
\begin{tabular}{ccccc}
\hline
\textbf{~~Version~~} &\textbf{~~Time~~}&\textbf{~~Constraints Size~~}&~~\#\textbf{.Div}~~&~~\#\textbf{.Opp}~~ \\ 
\hline
\texttt{Base} &~~~100\%~~~ &100\%    &12  &76\\ 
\texttt{Opt} &~~~30.7\%~~~ &78.9\%  &50 &79\\ 
\hline
\end{tabular}
}
\vspace{-2ex}
\end{table}

We also examine how the optimizations impact each benchmark in terms of 
the analysis time in Figure~\ref{fig:4a} and the constraint size in Figure
\ref{fig:4b}. Programs in these two figures are arranged using their orders 
in Table~\ref{tbl:benchs}. Tags \texttt{Base} and \texttt{Opt} also stand 
for the \SpecuSym versions without and with optimizations, respectively.

Clearly, Figure~\ref{fig:4a} shows that optimizations drastically reduce 
time cost for $6$ out of $15$ cases from hundreds of minutes to less than 
an hour. On the one hand, the \texttt{Base} has more repeated computations 
while \texttt{Opt} reuses cached intermediate results. On the other hand, 
formula reduction lowers the cost of solving. Figure~\ref{fig:4b} shows 
how optimizations reduce the constraint size where AES~\cite{LibTomCrypt}, 
glibc DES~\cite{glibc}, and seed~\cite{Tegra} have significant improvements. 
For the other cases, the memory trace length is commonly short. Thus, our 
optimizations have no apparent improvements upon them.

Finally, we answer the last research question: our optimizations significantly 
boost the overall performance by saving 69.3\% execution time and detecting 
31.8\% more timing leaks.

\subsection{Threat to Validity}
\label{sec:threat_validity}

\SpecuSym works on the LLVM IR. The program runtime behavior might differ due 
to compiler backend optimization and address space layout randomization. More 
low-level details can further facilitate \SpecuSym to improve the analysis.

\SpecuSym considers nested speculative execution because branch instruction 
may also exist in speculated paths. As the speculative window is usually 
transient, we treat the nested depth as configurable and set it to 2 in 
evaluation. However, experiments show the effectiveness of this 
under-approximation approach.

So far, \SpecuSym cannot support the analysis of multithreaded programs. Though 
{\sc{SymSC}}~\cite{GuoWW18} leveraged multithreading for timing leak detection, 
considering concurrency and speculative execution together greatly amplifies 
the state space and interleaving space, which pose exceptional challenges for a 
precise analysis.

Another threat is from the out-of-order execution, which schedules a set of 
data-independent instructions in an out-of-order manner, rather than sticking 
to the program order. This processor optimization may affect the accuracy of 
our analysis. However, incorporating both speculative execution and out-of-order 
execution requires more dedicated analysis. We leave it as future work.

\ignore{
We do not consider the situation that speculative execution leverages Branch 
Target Buffer (BTB) or Return Stack Buffer (RSB) to select the destinations 
of indirect branches. Such vulnerabilities have been effectively mitigated 
by Intel and AMD through Indirect Branch Prediction Barrier (IBPB) settings.
}

\section{Related Work}
\label{sec:related}

Regarding the security impact of processor speculative execution
\cite{kimuraKT1996}, Kocher et al.~\cite{KocherGGHHLMPSY19} demonstrated 
that speculative execution could silently influence the processor cache 
state, thus giving rise to cache timing attacks, which motivates the 
our \textsc{SpecuSym} work.

Symbolic execution~\cite{King76} has been applying to side-channel analysis 
because of the original designs of precise reasoning and input generation
\cite{PasareanuPM16,BangAPPB16,PhanBPMB17,BrennanSB18,BangRB18}. To quantify 
the information leakage through a cache timing side-channel, Chattopadhyay
et al.~\cite{ChattopadhyayBRZ17,Chattopadhyay17} and Basu et al.~\cite{BasuC17}
developed symbolic execution methods to study dependencies between sensitive 
data and cache behaviors. Wang et al. developed CacheD~\cite{WangWLZW17}, 
which targets timing leaks through a trace-based symbolic execution. 
Furthermore, Brotzman et al.~\cite{BrotzmanLZTK2018} proposed CaSym, a 
symbolic reasoning method that supports cache analysis upon multiple program 
paths. Considering the cache effects from thread interleavings, Guo et al.
\cite{GuoWW18} proposed \textsc{SymSC} to analyze cache timing leaks due to 
multithreading. These methods, however, all ignore cache changes due to 
speculative execution.

Abstract interpretation~\cite{CousotC77} has also been adopting to cache timing
side-channel analysis because of its scalable analysis and sound approximation. 
CacheAudit~\cite{DoychevFKMR13} approximates cache states by concrete state 
abstraction. Based on CacheAudit, Barthe et al.~\cite{BartheKMO14} track the 
program cache state affected by the attacker from a separate process. CacheS
\cite{WangBLWZW19} provides a secret-augmented abstract domain to track program 
secrets and dependencies. However, CacheAudit and CacheS are unaware of the 
impact of speculative execution. Though Wu et al.~\cite{WuW19} developed a 
dedicated abstract interpretation method for cache timing leak analysis under 
speculative execution, the proposed technique is incapable of generating precise 
speculative flow and inputs to diagnose leaks.

\ignore{
It treats public data in a coarse-grained 
fashion and secret data in a finer-grained manner to balance the scalability 
and precision.
}

Likewise, grey-box fuzz testing has been used in detecting program 
undefined behaviors~\cite{ChenLXGZZWL19}, Spectre-type vulnerabilities
\cite{OleksenkoTSF19}, exposing side channels~\cite{NilizadehNP19}, and 
revealing timing leaks~\cite{HeEC19}. However, they are either unaware 
of speculative execution or relying on sophisticated instrumentations on 
the control flow. Also, due to the fuzzing nature, these methods primarily 
count on various heuristics while \SpecuSym systematically solves for 
precise inputs.

Among the existing works, \textsc{KLEESpectre}~\cite{WangCBMR19}, 
\textsc{Spectector}~\cite{GuarnieriKMRS19}, and~\cite{ZhangCW12} are 
closely related to \textsc{SpecuSym}. \textsc{KLEESpectre}
\cite{WangCBMR19} has an analogous understanding of modeling speculation 
in symbolic execution. However, \textsc{KLEESpectre}~\cite{WangCBMR19} 
aims at Spectre-style vulnerability; thus, it has a crucial difference 
in constraint semantics. \textsc{Spectector}~\cite{WangCBMR19} also 
targets Spectre-type vulnerability. It employs symbolic execution to 
prove speculative non-interference property or detect violations. 
Speculative symbolic execution~\cite{ZhangCW12} proposed a method to 
lazily reason branch feasibility until reaching a threshold, which 
improves the efficiency of the symbolic execution technique.

Besides the above analysis methods, there also exists compiler assisted 
techniques~\cite{DoychevK17,WangS17,SungPW18,WuGSW18,WangSW19}, formal 
verifications 
\cite{EldibWS14,EldibWTS14,ChenFD17,SousaD16,AntoGHKTW17,ZhangGSW18}, 
and program synthesis approaches~\cite{MaffeiR17,EldibW14,EldibWW16}
that model or mitigate side-channel leaks. Still, none of them is 
applicable to cache timing side channels due to speculative execution.

\ignore{
Doychev et al.~\cite{DoychevK17}studied that compiler optimization can 
remove cache side channels. Sung et al.~\cite{SungPW18} developed a 
compiler frontend transformation that models cache timing behaviors 
for verification purposes.
}

To conclude, existing works may fail to produce inputs, ignore cache 
affects from speculative execution or leverages speculation for other 
analyses. In contrast, \SpecuSym models speculative execution through 
stateful symbolic execution to generate inputs and pinpoint leak sites, 
which can help reason the cache timing side-channel leaks caused by 
speculative execution in depth.

\section{Conclusions}
\label{sec:conclusion}

In this paper, we have presented a symbolic execution based method 
\SpecuSym for detecting cache timing leaks on running a sensitive data 
related program under speculative execution. \SpecuSym systematically 
explores the program state space and models speculative execution at 
conditional branches. By cache state modeling as well as cache behavior 
analysis, \SpecuSym leverages constraint-solving to search for divergent 
and opposite cache behaviors at memory access events. Experimental results
show that \SpecuSym successfully detects timing leaks under different cache 
settings. The proposed optimizations help \SpecuSym save 69.3\% execution 
time to detect 31.8\% more leaks. Comparison between \SpecuSym 
and state-of-the-art abstract interpretation based method~\cite{WuW19} shows 
that \SpecuSym not only successfully detects leaks in the reported programs 
but also eliminates false positives for a more precise result.

%
%

\clearpage\newpage
\bibliographystyle{ACM-Reference-Format}

\bibliography{tapsse}

\end{document}

%% file: fig_flow.tex
\definecolor{salmon}{RGB}{255,191,191}
\definecolor{leak}{RGB}{255,151,117}
\begin{tikzpicture}[fill=blue!20,font=\scriptsize] 
  \tikzstyle{arrow}=[thick,->,>=stealth,black]
  \tikzstyle{txt}=[above=5pt,right,text width=1.6cm]
  \tikzstyle{long_txt}=[above=5pt,right,text width=2cm]
  \tikzstyle{short_txt}=[above=5pt,right,text width=1.4cm]

  \draw [rounded corners=3] (0.0, 2.2) rectangle (1.5, 2.8);
  \node [above=5pt,right,text width=1.6cm] at (0.25, 2.45) {Symbolic};
  \node [above=5pt,right,text width=1.6cm] at (0.01, 2.2) {Sensitive Input};

  \draw [rounded corners=3] (0.0, 1.2) rectangle (1.5, 2.0);
  \node [above=5pt,right,text width=1.6cm] at (0.1, 1.55) {Leakage-free};
  \node [above=5pt,right,text width=1.6cm] at (0.15, 1.3) {Program $\prog$};

  \draw [rounded corners=3] (0.0, 0.4) rectangle (1.5, 1.0);
  \node [above=5pt,right,text width=1.6cm] at (0.17, 0.65) {Insensitive};
  \node [above=5pt,right,text width=1.6cm] at (0.15, 0.4) {Plain Input};

  \draw [arrow](1.5, 2.3)--(2.0, 1.7);
  \draw [arrow](1.5, 1.6)--(2.0, 1.6);
  \draw [arrow](1.5, 0.9)--(2.0, 1.5);

  \draw [fill=lightgray, rounded corners=3] (2.0, 1.2) rectangle (3.6, 2.0);
  \node [txt] at (2.32,1.55){Symbolic};
  \node [txt] at (2.3,1.3){Execution};

  \draw [fill=blue!20, rounded corners=3] (4.2, 0) rectangle (6.3, 0.8);
  \node [txt] at (4.9,0.35){Cache};
  \node [long_txt] at (4.5,0.1){State Modeling};
  \draw [arrow] (2.8,0.4)--(2.8,1.2);
  \draw (2.8,0.4)--(4.2,0.4);
  
  \draw [fill=blue!20, rounded corners=3](4.2, 1.2) rectangle (6.3, 2.0);
  \node [txt] at (4.9,1.55){Cache};
  \node [long_txt] at (4.35,1.3){Behavior Analysis};
  \draw [arrow](5.25,0.8)--(5.25,1.2);
  \draw [arrow](3.6,1.6)--(4.2,1.6);

  \draw [fill=blue!20, rounded corners=3] (4.2, 2.4) rectangle (6.3, 3.2);
  \node [txt] at (4.6, 2.75) {Speculative};
  \node [long_txt] at (4.3, 2.5) {Execution Modeling};
  \draw [arrow](2.8, 2.8)--(2.8, 2.0);
  \draw (2.8, 2.8)--(4.2, 2.8);
  \draw [arrow] (5.25, 2.4)--(5.25, 2.0);

  \draw [arrow](6.3,1.6)--(6.8,1.6);

  \draw [fill=orange, rounded corners=3](6.8, 1.2) rectangle (8.5, 2.0);
  \node [txt] at (7,1.55){New Timing};
  \node [txt] at (6.9,1.3){Leak Witnesses};

  \draw [dashed,black,thin,rounded corners=3](1.75, -0.2) rectangle (6.55, 3.4);

  \node [draw,circle,fill=white,minimum size=2pt,inner sep=2pt] at (4.2,3.18) {1};
  \node [draw,circle,fill=white,minimum size=2pt,inner sep=2pt] at (4.2,0.8) {2};
  \node [draw,circle,fill=white,minimum size=2pt,inner sep=2pt] at (4.2,2.0) {3};
  \node [draw,circle,fill=white,minimum size=2pt,inner sep=2pt] at (6.8,2.0) {4};

\end{tikzpicture}

%% file: tapsse.bbl

\begin{thebibliography}{00}


\ifx \showCODEN    \undefined \def \showCODEN     #1{\unskip}     \fi
\ifx \showDOI      \undefined \def \showDOI       #1{{\tt DOI:}\penalty0{#1}\ }
  \fi
\ifx \showISBNx    \undefined \def \showISBNx     #1{\unskip}     \fi
\ifx \showISBNxiii \undefined \def \showISBNxiii  #1{\unskip}     \fi
\ifx \showISSN     \undefined \def \showISSN      #1{\unskip}     \fi
\ifx \showLCCN     \undefined \def \showLCCN      #1{\unskip}     \fi
\ifx \shownote     \undefined \def \shownote      #1{#1}          \fi
\ifx \showarticletitle \undefined \def \showarticletitle #1{#1}   \fi
\ifx \showURL      \undefined \def \showURL       #1{#1}          \fi
\providecommand\bibfield[2]{#2}
\providecommand\bibinfo[2]{#2}
\providecommand\natexlab[1]{#1}
\providecommand\showeprint[2][]{arXiv:#2}

\bibitem[\protect\citeauthoryear{??}{gli}{}]%
        {glibc}
\bibinfo{booktitle}{{\em {glibc-2.29.9000}}}.
\newblock \url{https://www.gnu.org/software/libc/}.
\newblock


\bibitem[\protect\citeauthoryear{??}{hpn}{}]%
        {hpn-ssh}
\bibinfo{booktitle}{{\em {High Performance SSH/SCP - HPN-SSH}}}.
\newblock \url{https://www.psc.edu/hpn-ssh}.
\newblock


\bibitem[\protect\citeauthoryear{??}{Lib}{a}]%
        {Libgcrypt}
\bibinfo{booktitle}{{\em {Libgcrypt-1.8.4}}}.
\newblock \url{https://gnupg.org/software/libgcrypt/index.html}.
\newblock


\bibitem[\protect\citeauthoryear{??}{Lib}{b}]%
        {LibTomCrypt}
\bibinfo{booktitle}{{\em {LibTomCrypt}}}.
\newblock \url{http://www.libtom.net/LibTomCrypt/}.
\newblock


\bibitem[\protect\citeauthoryear{??}{Ope}{}]%
        {OpenSSL111c}
\bibinfo{booktitle}{{\em {OpenSSL-1.1.1c}}}.
\newblock
  \url{https://mta.openssl.org/pipermail/openssl-announce/2019-May/000153.html}.
\newblock


\bibitem[\protect\citeauthoryear{??}{spe}{}]%
        {spectre-attack}
\bibinfo{booktitle}{{\em {spectre-attach}}}.
\newblock \url{https://github.com/Eugnis/spectre-attack}.
\newblock


\bibitem[\protect\citeauthoryear{??}{Teg}{}]%
        {Tegra}
\bibinfo{booktitle}{{\em {Tegra}}}.
\newblock
  \url{https://android.googlesource.com/kernel/tegra/+/android-8.1.0_r0.113/crypto}.
\newblock


\bibitem[\protect\citeauthoryear{Antonopoulos, Gazzillo, Hicks, Koskinen,
  Terauchi, and Wei}{Antonopoulos et~al\mbox{.}}{2017}]%
        {AntoGHKTW17}
\bibfield{author}{\bibinfo{person}{Timos Antonopoulos}, \bibinfo{person}{Paul
  Gazzillo}, \bibinfo{person}{Michael Hicks}, \bibinfo{person}{Eric Koskinen},
  \bibinfo{person}{Tachio Terauchi}, {and} \bibinfo{person}{Shiyi Wei}.}
  \bibinfo{year}{2017}\natexlab{}.
\newblock \showarticletitle{Decomposition instead of self-composition for
  proving the absence of timing channels}. In \bibinfo{booktitle}{{\em
  Proceedings of the 38th {ACM} {SIGPLAN} Conference on Programming Language
  Design and Implementation, {PLDI} 2017, Barcelona, Spain, June 18-23, 2017}}.
  \bibinfo{pages}{362--375}.
\newblock


\bibitem[\protect\citeauthoryear{Bang, Aydin, Phan, Pasareanu, and Bultan}{Bang
  et~al\mbox{.}}{2016}]%
        {BangAPPB16}
\bibfield{author}{\bibinfo{person}{Lucas Bang}, \bibinfo{person}{Abdulbaki
  Aydin}, \bibinfo{person}{Quoc{-}Sang Phan}, \bibinfo{person}{Corina~S.
  Pasareanu}, {and} \bibinfo{person}{Tevfik Bultan}.}
  \bibinfo{year}{2016}\natexlab{}.
\newblock \showarticletitle{String analysis for side channels with segmented
  oracles}. In \bibinfo{booktitle}{{\em Proceedings of the 24th {ACM} {SIGSOFT}
  International Symposium on Foundations of Software Engineering, {FSE} 2016,
  Seattle, WA, USA, November 13-18, 2016}}. \bibinfo{pages}{193--204}.
\newblock


\bibitem[\protect\citeauthoryear{Bang, Rosner, and Bultan}{Bang
  et~al\mbox{.}}{2018}]%
        {BangRB18}
\bibfield{author}{\bibinfo{person}{Lucas Bang}, \bibinfo{person}{Nicol{\'{a}}s
  Rosner}, {and} \bibinfo{person}{Tevfik Bultan}.}
  \bibinfo{year}{2018}\natexlab{}.
\newblock \showarticletitle{Online Synthesis of Adaptive Side-Channel Attacks
  Based On Noisy Observations}. In \bibinfo{booktitle}{{\em 2018 {IEEE}
  European Symposium on Security and Privacy, EuroS{\&}P 2018, London, United
  Kingdom, April 24-26, 2018}}. \bibinfo{pages}{307--322}.
\newblock


\bibitem[\protect\citeauthoryear{Barthe, K{\"{o}}pf, Mauborgne, and
  Ochoa}{Barthe et~al\mbox{.}}{2014}]%
        {BartheKMO14}
\bibfield{author}{\bibinfo{person}{Gilles Barthe}, \bibinfo{person}{Boris
  K{\"{o}}pf}, \bibinfo{person}{Laurent Mauborgne}, {and}
  \bibinfo{person}{Mart{\'{\i}}n Ochoa}.} \bibinfo{year}{2014}\natexlab{}.
\newblock \showarticletitle{Leakage Resilience against Concurrent Cache
  Attacks}. In \bibinfo{booktitle}{{\em Principles of Security and Trust -
  Third International Conference, {POST} 2014}}. \bibinfo{pages}{140--158}.
\newblock


\bibitem[\protect\citeauthoryear{Basu and Chattopadhyay}{Basu and
  Chattopadhyay}{2017}]%
        {BasuC17}
\bibfield{author}{\bibinfo{person}{Tiyash Basu} {and} \bibinfo{person}{Sudipta
  Chattopadhyay}.} \bibinfo{year}{2017}\natexlab{}.
\newblock \showarticletitle{Testing Cache Side-Channel Leakage}. In
  \bibinfo{booktitle}{{\em 2017 {IEEE} International Conference on Software
  Testing, Verification and Validation Workshops, {ICST} Workshops 2017, Tokyo,
  Japan, March 13-17, 2017}}. \bibinfo{pages}{51--60}.
\newblock


\bibitem[\protect\citeauthoryear{Brennan, Saha, and Bultan}{Brennan
  et~al\mbox{.}}{2018}]%
        {BrennanSB18}
\bibfield{author}{\bibinfo{person}{Tegan Brennan}, \bibinfo{person}{Seemanta
  Saha}, {and} \bibinfo{person}{Tevfik Bultan}.}
  \bibinfo{year}{2018}\natexlab{}.
\newblock \showarticletitle{Symbolic path cost analysis for side-channel
  detection}. In \bibinfo{booktitle}{{\em Proceedings of the 40th International
  Conference on Software Engineering: Companion Proceeedings, {ICSE} 2018,
  Gothenburg, Sweden, May 27 - June 03, 2018}}. \bibinfo{pages}{424--425}.
\newblock


\bibitem[\protect\citeauthoryear{Brotzman, Liu, Zhang, Tan, and
  Kandemir}{Brotzman et~al\mbox{.}}{2019}]%
        {BrotzmanLZTK2018}
\bibfield{author}{\bibinfo{person}{Robert~L. Brotzman}, \bibinfo{person}{Shen
  Liu}, \bibinfo{person}{Danfeng Zhang}, \bibinfo{person}{Gang Tan}, {and}
  \bibinfo{person}{Mahmut~T. Kandemir}.} \bibinfo{year}{2019}\natexlab{}.
\newblock \showarticletitle{CaSym: Cache Aware Symbolic Execution for Side
  Channel Detection and Mitigation}. In \bibinfo{booktitle}{{\em IEEE Symposium
  on Security and Privacy}}.
\newblock


\bibitem[\protect\citeauthoryear{Bulck, Minkin, Weisse, Genkin, Kasikci,
  Piessens, Silberstein, Wenisch, Yarom, and Strackx}{Bulck
  et~al\mbox{.}}{2018}]%
        {BulckMWGKPSWYS18}
\bibfield{author}{\bibinfo{person}{Jo~Van Bulck}, \bibinfo{person}{Marina
  Minkin}, \bibinfo{person}{Ofir Weisse}, \bibinfo{person}{Daniel Genkin},
  \bibinfo{person}{Baris Kasikci}, \bibinfo{person}{Frank Piessens},
  \bibinfo{person}{Mark Silberstein}, \bibinfo{person}{Thomas~F. Wenisch},
  \bibinfo{person}{Yuval Yarom}, {and} \bibinfo{person}{Raoul Strackx}.}
  \bibinfo{year}{2018}\natexlab{}.
\newblock \showarticletitle{Foreshadow: Extracting the Keys to the Intel {SGX}
  Kingdom with Transient Out-of-Order Execution}. In \bibinfo{booktitle}{{\em
  27th {USENIX} Security Symposium, {USENIX} Security 2018, Baltimore, MD, USA,
  August 15-17, 2018.}} \bibinfo{pages}{991--1008}.
\newblock


\bibitem[\protect\citeauthoryear{C., Giri, and Menezes}{C.
  et~al\mbox{.}}{2016}]%
        {CGM16}
\bibfield{author}{\bibinfo{person}{Ashokkumar C.},
  \bibinfo{person}{Ravi~Prakash Giri}, {and} \bibinfo{person}{Bernard~L.
  Menezes}.} \bibinfo{year}{2016}\natexlab{}.
\newblock \showarticletitle{Highly Efficient Algorithms for {AES} Key Retrieval
  in Cache Access Attacks}. In \bibinfo{booktitle}{{\em {IEEE} European
  Symposium on Security and Privacy, EuroS{\&}P 2016, Saarbr{\"{u}}cken,
  Germany, March 21-24, 2016}}. \bibinfo{pages}{261--275}.
\newblock


\bibitem[\protect\citeauthoryear{Cadar, Dunbar, and Engler}{Cadar
  et~al\mbox{.}}{2008}]%
        {CadarDE08}
\bibfield{author}{\bibinfo{person}{Cristian Cadar}, \bibinfo{person}{Daniel
  Dunbar}, {and} \bibinfo{person}{Dawson~R. Engler}.}
  \bibinfo{year}{2008}\natexlab{}.
\newblock \showarticletitle{{KLEE:} Unassisted and Automatic Generation of
  High-Coverage Tests for Complex Systems Programs}. In
  \bibinfo{booktitle}{{\em 8th {USENIX} Symposium on Operating Systems Design
  and Implementation, {OSDI} 2008, December 8-10, 2008, San Diego, California,
  USA, Proceedings}}. \bibinfo{pages}{209--224}.
\newblock


\bibitem[\protect\citeauthoryear{Chattopadhyay}{Chattopadhyay}{2017}]%
        {Chattopadhyay17}
\bibfield{author}{\bibinfo{person}{Sudipta Chattopadhyay}.}
  \bibinfo{year}{2017}\natexlab{}.
\newblock \showarticletitle{Directed Automated Memory Performance Testing}. In
  \bibinfo{booktitle}{{\em Tools and Algorithms for the Construction and
  Analysis of Systems - 23rd International Conference, {TACAS} 2017, Held as
  Part of the European Joint Conferences on Theory and Practice of Software,
  {ETAPS}}}. \bibinfo{pages}{38--55}.
\newblock


\bibitem[\protect\citeauthoryear{Chattopadhyay, Beck, Rezine, and
  Zeller}{Chattopadhyay et~al\mbox{.}}{2017}]%
        {ChattopadhyayBRZ17}
\bibfield{author}{\bibinfo{person}{Sudipta Chattopadhyay},
  \bibinfo{person}{Moritz Beck}, \bibinfo{person}{Ahmed Rezine}, {and}
  \bibinfo{person}{Andreas Zeller}.} \bibinfo{year}{2017}\natexlab{}.
\newblock \showarticletitle{Quantifying the information leak in cache attacks
  via symbolic execution}. In \bibinfo{booktitle}{{\em Proceedings of the 15th
  {ACM-IEEE} International Conference on Formal Methods and Models for System
  Design, {MEMOCODE} 2017, Vienna, Austria, September 29 - October 02, 2017}}.
  \bibinfo{pages}{25--35}.
\newblock


\bibitem[\protect\citeauthoryear{Chen, Feng, and Dillig}{Chen
  et~al\mbox{.}}{2017}]%
        {ChenFD17}
\bibfield{author}{\bibinfo{person}{Jia Chen}, \bibinfo{person}{Yu Feng}, {and}
  \bibinfo{person}{Isil Dillig}.} \bibinfo{year}{2017}\natexlab{}.
\newblock \showarticletitle{Precise Detection of Side-Channel Vulnerabilities
  using Quantitative Cartesian Hoare Logic}. In \bibinfo{booktitle}{{\em
  Proceedings of the 2017 {ACM} {SIGSAC} Conference on Computer and
  Communications Security, {CCS} 2017, Dallas, TX, USA, October 30 - November
  03, 2017}}. \bibinfo{pages}{875--890}.
\newblock


\bibitem[\protect\citeauthoryear{Chen, Wang, Yang, and Stoller}{Chen
  et~al\mbox{.}}{2009}]%
        {ChenWYS09}
\bibfield{author}{\bibinfo{person}{Qichang Chen}, \bibinfo{person}{Liqiang
  Wang}, \bibinfo{person}{Zijiang Yang}, {and} \bibinfo{person}{Scott~D.
  Stoller}.} \bibinfo{year}{2009}\natexlab{}.
\newblock \showarticletitle{{HAVE:} Detecting Atomicity Violations via
  Integrated Dynamic and Static Analysis}. In \bibinfo{booktitle}{{\em
  Fundamental Approaches to Software Engineering, 12th International
  Conference, {FASE} 2009, Held as Part of the Joint European Conferences on
  Theory and Practice of Software, {ETAPS} 2009, York, UK, March 22-29, 2009.
  Proceedings}}. \bibinfo{pages}{425--439}.
\newblock


\bibitem[\protect\citeauthoryear{Chen, Lin, Dai, Hsu, and Yew}{Chen
  et~al\mbox{.}}{2004}]%
        {ChenLDHY04}
\bibfield{author}{\bibinfo{person}{Tong Chen}, \bibinfo{person}{Jin Lin},
  \bibinfo{person}{Xiaoru Dai}, \bibinfo{person}{Wei{-}Chung Hsu}, {and}
  \bibinfo{person}{Pen{-}Chung Yew}.} \bibinfo{year}{2004}\natexlab{}.
\newblock \showarticletitle{Data Dependence Profiling for Speculative
  Optimizations}. In \bibinfo{booktitle}{{\em Compiler Construction, 13th
  International Conference, {CC} 2004, Held as Part of the Joint European
  Conferences on Theory and Practice of Software, {ETAPS} 2004, Barcelona,
  Spain, March 29 - April 2, 2004, Proceedings}}. \bibinfo{pages}{57--72}.
\newblock


\bibitem[\protect\citeauthoryear{Chen, Li, Xu, Guo, Zhou, Zhang, Wei, and
  Lu}{Chen et~al\mbox{.}}{2019}]%
        {ChenLXGZZWL19}
\bibfield{author}{\bibinfo{person}{Yaohui Chen}, \bibinfo{person}{Peng Li},
  \bibinfo{person}{Jun Xu}, \bibinfo{person}{Shengjian Guo},
  \bibinfo{person}{Rundong Zhou}, \bibinfo{person}{Yulong Zhang},
  \bibinfo{person}{Tao Wei}, {and} \bibinfo{person}{Long Lu}.}
  \bibinfo{year}{2019}\natexlab{}.
\newblock \showarticletitle{{SAVIOR:} Towards Bug-Driven Hybrid Testing}.
\newblock \bibinfo{journal}{{\em CoRR\/}}  \bibinfo{volume}{abs/1906.07327}
  (\bibinfo{year}{2019}).
\newblock


\bibitem[\protect\citeauthoryear{Clarke}{Clarke}{1976}]%
        {Clarke76}
\bibfield{author}{\bibinfo{person}{Lori~A. Clarke}.}
  \bibinfo{year}{1976}\natexlab{}.
\newblock \showarticletitle{A program testing system}. In
  \bibinfo{booktitle}{{\em Proceedings of the 1976 Annual Conference, Houston,
  Texas, USA, October 20-22, 1976}}. \bibinfo{pages}{488--491}.
\newblock


\bibitem[\protect\citeauthoryear{Cousot and Cousot}{Cousot and Cousot}{1977}]%
        {CousotC77}
\bibfield{author}{\bibinfo{person}{Patrick Cousot} {and}
  \bibinfo{person}{Radhia Cousot}.} \bibinfo{year}{1977}\natexlab{}.
\newblock \showarticletitle{Abstract Interpretation: {A} Unified Lattice Model
  for Static Analysis of Programs by Construction or Approximation of
  Fixpoints}. In \bibinfo{booktitle}{{\em Conference Record of the Fourth {ACM}
  Symposium on Principles of Programming Languages, Los Angeles, California,
  USA, January 1977}}. \bibinfo{pages}{238--252}.
\newblock


\bibitem[\protect\citeauthoryear{Dhem, Koeune, Leroux, Mestr{\'{e}},
  Quisquater, and Willems}{Dhem et~al\mbox{.}}{1998}]%
        {DhemKLMQW98}
\bibfield{author}{\bibinfo{person}{Jean{-}Fran{\c{c}}ois Dhem},
  \bibinfo{person}{Fran{\c{c}}ois Koeune},
  \bibinfo{person}{Philippe{-}Alexandre Leroux}, \bibinfo{person}{Patrick
  Mestr{\'{e}}}, \bibinfo{person}{Jean{-}Jacques Quisquater}, {and}
  \bibinfo{person}{Jean{-}Louis Willems}.} \bibinfo{year}{1998}\natexlab{}.
\newblock \showarticletitle{A Practical Implementation of the Timing Attack}.
  In \bibinfo{booktitle}{{\em Smart Card Research and Applications, This
  International Conference, {CARDIS} '98, Louvain-la-Neuve, Belgium, September
  14-16, 1998, Proceedings}}. \bibinfo{pages}{167--182}.
\newblock


\bibitem[\protect\citeauthoryear{Disselkoen, Kohlbrenner, Porter, and
  Tullsen}{Disselkoen et~al\mbox{.}}{2017}]%
        {DisselkoenKPT17}
\bibfield{author}{\bibinfo{person}{Craig Disselkoen}, \bibinfo{person}{David
  Kohlbrenner}, \bibinfo{person}{Leo Porter}, {and} \bibinfo{person}{Dean~M.
  Tullsen}.} \bibinfo{year}{2017}\natexlab{}.
\newblock \showarticletitle{Prime+Abort: {A} Timer-Free High-Precision {L3}
  Cache Attack using Intel {TSX}}. In \bibinfo{booktitle}{{\em 26th {USENIX}
  Security Symposium, {USENIX} Security 2017, Vancouver, BC, Canada, August
  16-18, 2017.}} \bibinfo{pages}{51--67}.
\newblock


\bibitem[\protect\citeauthoryear{Doychev, Feld, K{\"{o}}pf, Mauborgne, and
  Reineke}{Doychev et~al\mbox{.}}{2013}]%
        {DoychevFKMR13}
\bibfield{author}{\bibinfo{person}{Goran Doychev}, \bibinfo{person}{Dominik
  Feld}, \bibinfo{person}{Boris K{\"{o}}pf}, \bibinfo{person}{Laurent
  Mauborgne}, {and} \bibinfo{person}{Jan Reineke}.}
  \bibinfo{year}{2013}\natexlab{}.
\newblock \showarticletitle{CacheAudit: {A} Tool for the Static Analysis of
  Cache Side Channels}. In \bibinfo{booktitle}{{\em Proceedings of the 22th
  {USENIX} Security Symposium, Washington, DC, USA, August 14-16, 2013}}.
  \bibinfo{pages}{431--446}.
\newblock


\bibitem[\protect\citeauthoryear{Doychev and K{\"{o}}pf}{Doychev and
  K{\"{o}}pf}{2017}]%
        {DoychevK17}
\bibfield{author}{\bibinfo{person}{Goran Doychev} {and} \bibinfo{person}{Boris
  K{\"{o}}pf}.} \bibinfo{year}{2017}\natexlab{}.
\newblock \showarticletitle{Rigorous analysis of software countermeasures
  against cache attacks}. In \bibinfo{booktitle}{{\em Proceedings of the 38th
  {ACM} {SIGPLAN} Conference on Programming Language Design and Implementation,
  {PLDI} 2017, Barcelona, Spain, June 18-23, 2017}}. \bibinfo{pages}{406--421}.
\newblock


\bibitem[\protect\citeauthoryear{Duddu, Samanta, Rao, and Balas}{Duddu
  et~al\mbox{.}}{2018}]%
        {DudduSRB18}
\bibfield{author}{\bibinfo{person}{Vasisht Duddu}, \bibinfo{person}{Debasis
  Samanta}, \bibinfo{person}{D.~Vijay Rao}, {and} \bibinfo{person}{Valentina~E.
  Balas}.} \bibinfo{year}{2018}\natexlab{}.
\newblock \showarticletitle{Stealing Neural Networks via Timing Side Channels}.
\newblock \bibinfo{journal}{{\em CoRR\/}}  \bibinfo{volume}{abs/1812.11720}
  (\bibinfo{year}{2018}).
\newblock
\showeprint[arxiv]{1812.11720}
\showURL{%
\url{http://arxiv.org/abs/1812.11720}}


\bibitem[\protect\citeauthoryear{Eldib and Wang}{Eldib and Wang}{2014}]%
        {EldibW14}
\bibfield{author}{\bibinfo{person}{Hassan Eldib} {and} \bibinfo{person}{Chao
  Wang}.} \bibinfo{year}{2014}\natexlab{}.
\newblock \showarticletitle{Synthesis of Masking Countermeasures against Side
  Channel Attacks}. In \bibinfo{booktitle}{{\em Computer Aided Verification -
  26th International Conference, {CAV} 2014, Held as Part of the Vienna Summer
  of Logic, {VSL} 2014, Vienna, Austria, July 18-22, 2014. Proceedings}}.
  \bibinfo{pages}{114--130}.
\newblock


\bibitem[\protect\citeauthoryear{Eldib, Wang, and Schaumont}{Eldib
  et~al\mbox{.}}{2014a}]%
        {EldibWS14}
\bibfield{author}{\bibinfo{person}{Hassan Eldib}, \bibinfo{person}{Chao Wang},
  {and} \bibinfo{person}{Patrick Schaumont}.} \bibinfo{year}{2014}\natexlab{a}.
\newblock \showarticletitle{SMT-Based Verification of Software Countermeasures
  against Side-Channel Attacks}. In \bibinfo{booktitle}{{\em Tools and
  Algorithms for the Construction and Analysis of Systems - 20th International
  Conference, {TACAS} 2014, Held as Part of the European Joint Conferences on
  Theory and Practice of Software, {ETAPS} 2014, Grenoble, France, April 5-13,
  2014. Proceedings}}. \bibinfo{pages}{62--77}.
\newblock


\bibitem[\protect\citeauthoryear{Eldib, Wang, Taha, and Schaumont}{Eldib
  et~al\mbox{.}}{2014b}]%
        {EldibWTS14}
\bibfield{author}{\bibinfo{person}{Hassan Eldib}, \bibinfo{person}{Chao Wang},
  \bibinfo{person}{Mostafa M.~I. Taha}, {and} \bibinfo{person}{Patrick
  Schaumont}.} \bibinfo{year}{2014}\natexlab{b}.
\newblock \showarticletitle{{QMS:} Evaluating the Side-Channel Resistance of
  Masked Software from Source Code}. In \bibinfo{booktitle}{{\em The 51st
  Annual Design Automation Conference 2014, {DAC} '14, San Francisco, CA, USA,
  June 1-5, 2014}}. \bibinfo{pages}{209:1--209:6}.
\newblock


\bibitem[\protect\citeauthoryear{Eldib, Wu, and Wang}{Eldib
  et~al\mbox{.}}{2016}]%
        {EldibWW16}
\bibfield{author}{\bibinfo{person}{Hassan Eldib}, \bibinfo{person}{Meng Wu},
  {and} \bibinfo{person}{Chao Wang}.} \bibinfo{year}{2016}\natexlab{}.
\newblock \showarticletitle{Synthesis of Fault-Attack Countermeasures for
  Cryptographic Circuits}. In \bibinfo{booktitle}{{\em Computer Aided
  Verification - 28th International Conference, {CAV} 2016, Toronto, ON,
  Canada, July 17-23, 2016, Proceedings, Part {II}}}.
  \bibinfo{pages}{343--363}.
\newblock


\bibitem[\protect\citeauthoryear{Guarnieri, K{\"{o}}pf, Morales, Reineke, and
  S{\'{a}}nchez}{Guarnieri et~al\mbox{.}}{2018}]%
        {GuarnieriKMRS19}
\bibfield{author}{\bibinfo{person}{Marco Guarnieri}, \bibinfo{person}{Boris
  K{\"{o}}pf}, \bibinfo{person}{Jos{\'{e}}~F. Morales}, \bibinfo{person}{Jan
  Reineke}, {and} \bibinfo{person}{Andr{\'{e}}s S{\'{a}}nchez}.}
  \bibinfo{year}{2018}\natexlab{}.
\newblock \showarticletitle{{Spectector:} Principled Detection of Speculative
  Information Flows}.
\newblock \bibinfo{journal}{{\em CoRR\/}}  \bibinfo{volume}{abs/1812.08639}
  (\bibinfo{year}{2018}).
\newblock
\showeprint[arxiv]{1812.08639}
\showURL{%
\url{http://arxiv.org/abs/1812.08639}}


\bibitem[\protect\citeauthoryear{Gullasch, Bangerter, and Krenn}{Gullasch
  et~al\mbox{.}}{2011}]%
        {GullaschBK11}
\bibfield{author}{\bibinfo{person}{David Gullasch}, \bibinfo{person}{Endre
  Bangerter}, {and} \bibinfo{person}{Stephan Krenn}.}
  \bibinfo{year}{2011}\natexlab{}.
\newblock \showarticletitle{Cache Games - Bringing Access-Based Cache Attacks
  on {AES} to Practice}. In \bibinfo{booktitle}{{\em 32nd {IEEE} Symposium on
  Security and Privacy, S{\&}P 2011, 22-25 May 2011, Berkeley, California,
  {USA}}}. \bibinfo{pages}{490--505}.
\newblock


\bibitem[\protect\citeauthoryear{Guo, Kusano, Wang, Yang, and Gupta}{Guo
  et~al\mbox{.}}{2015}]%
        {GuoKWYG15}
\bibfield{author}{\bibinfo{person}{Shengjian Guo}, \bibinfo{person}{Markus
  Kusano}, \bibinfo{person}{Chao Wang}, \bibinfo{person}{Zijiang Yang}, {and}
  \bibinfo{person}{Aarti Gupta}.} \bibinfo{year}{2015}\natexlab{}.
\newblock \showarticletitle{Assertion guided symbolic execution of
  multithreaded programs}. In \bibinfo{booktitle}{{\em ACM SIGSOFT Symposium on
  Foundations of Software Engineering}}. \bibinfo{pages}{854--865}.
\newblock


\bibitem[\protect\citeauthoryear{Guo, Wu, and Wang}{Guo et~al\mbox{.}}{2018}]%
        {GuoWW18}
\bibfield{author}{\bibinfo{person}{Shengjian Guo}, \bibinfo{person}{Meng Wu},
  {and} \bibinfo{person}{Chao Wang}.} \bibinfo{year}{2018}\natexlab{}.
\newblock \showarticletitle{Adversarial symbolic execution for detecting
  concurrency-related cache timing leaks}. In \bibinfo{booktitle}{{\em
  Proceedings of the 2018 {ACM} Joint Meeting on European Software Engineering
  Conference and Symposium on the Foundations of Software Engineering,
  {ESEC/SIGSOFT} {FSE} 2018, Lake Buena Vista, FL, USA, November 04-09, 2018}}.
  \bibinfo{pages}{377--388}.
\newblock


\bibitem[\protect\citeauthoryear{He, Emmi, and Ciocarlie}{He
  et~al\mbox{.}}{2019}]%
        {HeEC19}
\bibfield{author}{\bibinfo{person}{Shaobo He}, \bibinfo{person}{Michael Emmi},
  {and} \bibinfo{person}{Gabriela~F. Ciocarlie}.}
  \bibinfo{year}{2019}\natexlab{}.
\newblock \showarticletitle{ct-fuzz: Fuzzing for Timing Leaks}.
\newblock \bibinfo{journal}{{\em CoRR\/}}  \bibinfo{volume}{abs/1904.07280}
  (\bibinfo{year}{2019}).
\newblock


\bibitem[\protect\citeauthoryear{Hong, Davinroy, Kaya, Locke, Rackow, Kulda,
  Dachman{-}Soled, and Dumitras}{Hong et~al\mbox{.}}{2018}]%
        {HongDKLRKDD18}
\bibfield{author}{\bibinfo{person}{Sanghyun Hong}, \bibinfo{person}{Michael
  Davinroy}, \bibinfo{person}{Yigitcan Kaya}, \bibinfo{person}{Stuart~Nevans
  Locke}, \bibinfo{person}{Ian Rackow}, \bibinfo{person}{Kevin Kulda},
  \bibinfo{person}{Dana Dachman{-}Soled}, {and} \bibinfo{person}{Tudor
  Dumitras}.} \bibinfo{year}{2018}\natexlab{}.
\newblock \showarticletitle{Security Analysis of Deep Neural Networks Operating
  in the Presence of Cache Side-Channel Attacks}.
\newblock \bibinfo{journal}{{\em CoRR\/}}  \bibinfo{volume}{abs/1810.03487}
  (\bibinfo{year}{2018}).
\newblock
\showeprint[arxiv]{1810.03487}
\showURL{%
\url{http://arxiv.org/abs/1810.03487}}


\bibitem[\protect\citeauthoryear{Hu, Liang, Deng, Li, Xie, Ji, Ding, Liu,
  Sherwood, and Xie}{Hu et~al\mbox{.}}{2019}]%
        {HuLDLXJDLSX18}
\bibfield{author}{\bibinfo{person}{Xing Hu}, \bibinfo{person}{Ling Liang},
  \bibinfo{person}{Lei Deng}, \bibinfo{person}{Shuangchen Li},
  \bibinfo{person}{Xinfeng Xie}, \bibinfo{person}{Yu Ji},
  \bibinfo{person}{Yufei Ding}, \bibinfo{person}{Chang Liu},
  \bibinfo{person}{Timothy Sherwood}, {and} \bibinfo{person}{Yuan Xie}.}
  \bibinfo{year}{2019}\natexlab{}.
\newblock \showarticletitle{Neural Network Model Extraction Attacks in Edge
  Devices by Hearing Architectural Hints}.
\newblock \bibinfo{journal}{{\em CoRR\/}}  \bibinfo{volume}{abs/1903.03916}
  (\bibinfo{year}{2019}).
\newblock
\showeprint[arxiv]{1903.03916}
\showURL{%
\url{http://arxiv.org/abs/1903.03916}}


\bibitem[\protect\citeauthoryear{Hund, Willems, and Holz}{Hund
  et~al\mbox{.}}{2013}]%
        {HundWH13}
\bibfield{author}{\bibinfo{person}{Ralf Hund}, \bibinfo{person}{Carsten
  Willems}, {and} \bibinfo{person}{Thorsten Holz}.}
  \bibinfo{year}{2013}\natexlab{}.
\newblock \showarticletitle{Practical Timing Side Channel Attacks against
  Kernel Space {ASLR}}. In \bibinfo{booktitle}{{\em 2013 {IEEE} Symposium on
  Security and Privacy, {SP} 2013, Berkeley, CA, USA, May 19-22, 2013}}.
  \bibinfo{pages}{191--205}.
\newblock


\bibitem[\protect\citeauthoryear{Islam, Moghimi, Bruhns, Krebbel,
  G{\"{u}}lmezoglu, Eisenbarth, and Sunar}{Islam et~al\mbox{.}}{2019}]%
        {IslamMBKGES19}
\bibfield{author}{\bibinfo{person}{Saad Islam}, \bibinfo{person}{Ahmad
  Moghimi}, \bibinfo{person}{Ida Bruhns}, \bibinfo{person}{Moritz Krebbel},
  \bibinfo{person}{Berk G{\"{u}}lmezoglu}, \bibinfo{person}{Thomas Eisenbarth},
  {and} \bibinfo{person}{Berk Sunar}.} \bibinfo{year}{2019}\natexlab{}.
\newblock \showarticletitle{{SPOILER:} Speculative Load Hazards Boost Rowhammer
  and Cache Attacks}.
\newblock \bibinfo{journal}{{\em CoRR\/}}  \bibinfo{volume}{abs/1903.00446}
  (\bibinfo{year}{2019}).
\newblock
\showeprint[arxiv]{1903.00446}
\showURL{%
\url{http://arxiv.org/abs/1903.00446}}


\bibitem[\protect\citeauthoryear{Kimura, Yoshioka, and Kiyohara}{Kimura
  et~al\mbox{.}}{1996}]%
        {kimuraKT1996}
\bibfield{author}{\bibinfo{person}{Kozo Kimura}, \bibinfo{person}{Kosuki
  Yoshioka}, {and} \bibinfo{person}{Tokuzo Kiyohara}.}
  \bibinfo{year}{1996}\natexlab{}.
\newblock \bibinfo{title}{Speculative execution processor}.
\newblock   (\bibinfo{date}{April~23} \bibinfo{year}{1996}).
\newblock
\newblock
\shownote{US Patent 5,511,172.}


\bibitem[\protect\citeauthoryear{King}{King}{1976}]%
        {King76}
\bibfield{author}{\bibinfo{person}{James~C. King}.}
  \bibinfo{year}{1976}\natexlab{}.
\newblock \showarticletitle{Symbolic Execution and Program Testing}.
\newblock \bibinfo{journal}{{\em Commun. {ACM}\/}} \bibinfo{volume}{19},
  \bibinfo{number}{7} (\bibinfo{year}{1976}), \bibinfo{pages}{385--394}.
\newblock


\bibitem[\protect\citeauthoryear{Kocher, Horn, Fogh, Genkin, Gruss, Haas,
  Hamburg, Lipp, Mangard, Prescher, Schwarz, and Yarom}{Kocher
  et~al\mbox{.}}{2019}]%
        {KocherGGHHLMPSY19}
\bibfield{author}{\bibinfo{person}{Paul Kocher}, \bibinfo{person}{Jann Horn},
  \bibinfo{person}{Anders Fogh}, \bibinfo{person}{Daniel Genkin},
  \bibinfo{person}{Daniel Gruss}, \bibinfo{person}{Werner Haas},
  \bibinfo{person}{Mike Hamburg}, \bibinfo{person}{Moritz Lipp},
  \bibinfo{person}{Stefan Mangard}, \bibinfo{person}{Thomas Prescher},
  \bibinfo{person}{Michael Schwarz}, {and} \bibinfo{person}{Yuval Yarom}.}
  \bibinfo{year}{2019}\natexlab{}.
\newblock \showarticletitle{Spectre Attacks: Exploiting Speculative Execution}.
  In \bibinfo{booktitle}{{\em 40th IEEE Symposium on Security and Privacy
  (S\&P'19)}}.
\newblock


\bibitem[\protect\citeauthoryear{Kocher}{Kocher}{1996}]%
        {Kocher96}
\bibfield{author}{\bibinfo{person}{Paul~C. Kocher}.}
  \bibinfo{year}{1996}\natexlab{}.
\newblock \showarticletitle{Timing Attacks on Implementations of
  Diffie-Hellman, RSA, DSS, and Other Systems}. In \bibinfo{booktitle}{{\em
  Advances in Cryptology - {CRYPTO} '96, 16th Annual International Cryptology
  Conference, Santa Barbara, California, USA, August 18-22, 1996,
  Proceedings}}. \bibinfo{pages}{104--113}.
\newblock


\bibitem[\protect\citeauthoryear{Lattner and Adve}{Lattner and Adve}{2004}]%
        {LattnerA04}
\bibfield{author}{\bibinfo{person}{Chris Lattner} {and}
  \bibinfo{person}{Vikram~S. Adve}.} \bibinfo{year}{2004}\natexlab{}.
\newblock \showarticletitle{{LLVM:} {A} Compilation Framework for Lifelong
  Program Analysis {\&} Transformation}. In \bibinfo{booktitle}{{\em 2nd
  {IEEE}/{ACM} International Symposium on Code Generation and Optimization,
  20-24 March 2004, San Jose, CA, {USA}}}. \bibinfo{pages}{75--88}.
\newblock


\bibitem[\protect\citeauthoryear{Li, Ellis, Lebeck, and Sorin}{Li
  et~al\mbox{.}}{2005}]%
        {LiELS05}
\bibfield{author}{\bibinfo{person}{Tong Li}, \bibinfo{person}{Carla~Schlatter
  Ellis}, \bibinfo{person}{Alvin~R. Lebeck}, {and} \bibinfo{person}{Daniel~J.
  Sorin}.} \bibinfo{year}{2005}\natexlab{}.
\newblock \showarticletitle{Pulse: {A} Dynamic Deadlock Detection Mechanism
  Using Speculative Execution}. In \bibinfo{booktitle}{{\em Proceedings of the
  2005 {USENIX} Annual Technical Conference, April 10-15, 2005, Anaheim, CA,
  {USA}}}. \bibinfo{pages}{31--44}.
\newblock


\bibitem[\protect\citeauthoryear{Li, Mitra, and Roychoudhury}{Li
  et~al\mbox{.}}{2003}]%
        {LiMR03}
\bibfield{author}{\bibinfo{person}{Xianfeng Li}, \bibinfo{person}{Tulika
  Mitra}, {and} \bibinfo{person}{Abhik Roychoudhury}.}
  \bibinfo{year}{2003}\natexlab{}.
\newblock \showarticletitle{Accurate timing analysis by modeling caches,
  speculation and their interaction}. In \bibinfo{booktitle}{{\em Proceedings
  of the 40th Design Automation Conference, {DAC} 2003, Anaheim, CA, USA, June
  2-6, 2003}}. \bibinfo{pages}{466--471}.
\newblock


\bibitem[\protect\citeauthoryear{Li, Mitra, and Roychoudhury}{Li
  et~al\mbox{.}}{2005}]%
        {LiMR05}
\bibfield{author}{\bibinfo{person}{Xianfeng Li}, \bibinfo{person}{Tulika
  Mitra}, {and} \bibinfo{person}{Abhik Roychoudhury}.}
  \bibinfo{year}{2005}\natexlab{}.
\newblock \showarticletitle{Modeling Control Speculation for Timing Analysis}.
\newblock \bibinfo{journal}{{\em Real-Time Systems\/}} \bibinfo{volume}{29},
  \bibinfo{number}{1} (\bibinfo{year}{2005}), \bibinfo{pages}{27--58}.
\newblock


\bibitem[\protect\citeauthoryear{Lipp, Schwarz, Gruss, Prescher, Haas, Fogh,
  Horn, Mangard, Kocher, Genkin, Yarom, and Hamburg}{Lipp
  et~al\mbox{.}}{2018}]%
        {LippSGPHFHMKGYH18}
\bibfield{author}{\bibinfo{person}{Moritz Lipp}, \bibinfo{person}{Michael
  Schwarz}, \bibinfo{person}{Daniel Gruss}, \bibinfo{person}{Thomas Prescher},
  \bibinfo{person}{Werner Haas}, \bibinfo{person}{Anders Fogh},
  \bibinfo{person}{Jann Horn}, \bibinfo{person}{Stefan Mangard},
  \bibinfo{person}{Paul Kocher}, \bibinfo{person}{Daniel Genkin},
  \bibinfo{person}{Yuval Yarom}, {and} \bibinfo{person}{Mike Hamburg}.}
  \bibinfo{year}{2018}\natexlab{}.
\newblock \showarticletitle{Meltdown: Reading Kernel Memory from User Space}.
  In \bibinfo{booktitle}{{\em 27th {USENIX} Security Symposium, {USENIX}
  Security 2018, Baltimore, MD, USA, August 15-17, 2018.}}
  \bibinfo{pages}{973--990}.
\newblock


\bibitem[\protect\citeauthoryear{Maffei and Ryan}{Maffei and Ryan}{2017}]%
        {MaffeiR17}
\bibfield{editor}{\bibinfo{person}{Matteo Maffei} {and} \bibinfo{person}{Mark
  Ryan}} (Eds.). \bibinfo{year}{2017}\natexlab{}.
\newblock \bibinfo{booktitle}{{\em Principles of Security and Trust - 6th
  International Conference, {POST} 2017, Held as Part of the European Joint
  Conferences on Theory and Practice of Software, {ETAPS} 2017, Uppsala,
  Sweden, April 22-29, 2017, Proceedings}}. \bibinfo{series}{Lecture Notes in
  Computer Science}, Vol.~\bibinfo{volume}{10204}.
  \bibinfo{publisher}{Springer}.
\newblock


\bibitem[\protect\citeauthoryear{Mittal}{Mittal}{2019}]%
        {Mittal19}
\bibfield{author}{\bibinfo{person}{Sparsh Mittal}.}
  \bibinfo{year}{2019}\natexlab{}.
\newblock \showarticletitle{A survey of techniques for dynamic branch
  prediction}.
\newblock \bibinfo{journal}{{\em Concurrency and Computation: Practice and
  Experience\/}} \bibinfo{volume}{31}, \bibinfo{number}{1}
  (\bibinfo{year}{2019}).
\newblock


\bibitem[\protect\citeauthoryear{Moshovos and Sohi}{Moshovos and Sohi}{1997}]%
        {MoshovosS97}
\bibfield{author}{\bibinfo{person}{Andreas Moshovos} {and}
  \bibinfo{person}{Gurindar~S. Sohi}.} \bibinfo{year}{1997}\natexlab{}.
\newblock \showarticletitle{Streamlining Inter-Operation Memory Communication
  via Data Dependence Prediction}. In \bibinfo{booktitle}{{\em Proceedings of
  the Thirtieth Annual {IEEE/ACM} International Symposium on Microarchitecture,
  {MICRO} 30, Research Triangle Park, North Carolina, USA, December 1-3,
  1997}}. \bibinfo{pages}{235--245}.
\newblock


\bibitem[\protect\citeauthoryear{Nicolau}{Nicolau}{1989}]%
        {Nicolau89}
\bibfield{author}{\bibinfo{person}{Alexandru Nicolau}.}
  \bibinfo{year}{1989}\natexlab{}.
\newblock \showarticletitle{Run-Time Disambiguation: Coping with Statically
  Unpredictable Dependencies}.
\newblock \bibinfo{journal}{{\em {IEEE} Trans. Computers\/}}
  \bibinfo{volume}{38}, \bibinfo{number}{5} (\bibinfo{year}{1989}),
  \bibinfo{pages}{663--678}.
\newblock


\bibitem[\protect\citeauthoryear{Nilizadeh, Noller, and Pasareanu}{Nilizadeh
  et~al\mbox{.}}{2019}]%
        {NilizadehNP19}
\bibfield{author}{\bibinfo{person}{Shirin Nilizadeh}, \bibinfo{person}{Yannic
  Noller}, {and} \bibinfo{person}{Corina~S. Pasareanu}.}
  \bibinfo{year}{2019}\natexlab{}.
\newblock \showarticletitle{DifFuzz: differential fuzzing for side-channel
  analysis}. In \bibinfo{booktitle}{{\em Proceedings of the 41st International
  Conference on Software Engineering, {ICSE} 2019, Montreal, QC, Canada, May
  25-31, 2019}}. \bibinfo{pages}{176--187}.
\newblock


\bibitem[\protect\citeauthoryear{Oleksenko, Trach, Silberstein, and
  Fetzer}{Oleksenko et~al\mbox{.}}{2019}]%
        {OleksenkoTSF19}
\bibfield{author}{\bibinfo{person}{Oleksii Oleksenko}, \bibinfo{person}{Bohdan
  Trach}, \bibinfo{person}{Mark Silberstein}, {and} \bibinfo{person}{Christof
  Fetzer}.} \bibinfo{year}{2019}\natexlab{}.
\newblock \showarticletitle{SpecFuzz: Bringing Spectre-type vulnerabilities to
  the surface}.
\newblock \bibinfo{journal}{{\em CoRR\/}}  \bibinfo{volume}{abs/1905.10311}
  (\bibinfo{year}{2019}).
\newblock
\showeprint[arxiv]{1905.10311}
\showURL{%
\url{http://arxiv.org/abs/1905.10311}}


\bibitem[\protect\citeauthoryear{{\"{O}}nder and Gupta}{{\"{O}}nder and
  Gupta}{1999}]%
        {OnderG99}
\bibfield{author}{\bibinfo{person}{Soner {\"{O}}nder} {and}
  \bibinfo{person}{Rajiv Gupta}.} \bibinfo{year}{1999}\natexlab{}.
\newblock \showarticletitle{Dynamic Memory Disambiguation in the Presence of
  Out-of-Order Store Issuing}. In \bibinfo{booktitle}{{\em Proceedings of the
  32nd Annual {IEEE/ACM} International Symposium on Microarchitecture, {MICRO}
  32, Haifa, Israel, November 16-18, 1999}}. \bibinfo{pages}{170--176}.
\newblock


\bibitem[\protect\citeauthoryear{Osvik, Shamir, and Tromer}{Osvik
  et~al\mbox{.}}{2006}]%
        {OsvikST06}
\bibfield{author}{\bibinfo{person}{Dag~Arne Osvik}, \bibinfo{person}{Adi
  Shamir}, {and} \bibinfo{person}{Eran Tromer}.}
  \bibinfo{year}{2006}\natexlab{}.
\newblock \showarticletitle{Cache Attacks and Countermeasures: The Case of
  {AES}}. In \bibinfo{booktitle}{{\em Topics in Cryptology - {CT-RSA} 2006, The
  Cryptographers' Track at the {RSA} Conference 2006, San Jose, CA, USA,
  February 13-17, 2006, Proceedings}}. \bibinfo{pages}{1--20}.
\newblock


\bibitem[\protect\citeauthoryear{Pasareanu, Phan, and Malacaria}{Pasareanu
  et~al\mbox{.}}{2016}]%
        {PasareanuPM16}
\bibfield{author}{\bibinfo{person}{Corina~S. Pasareanu},
  \bibinfo{person}{Quoc{-}Sang Phan}, {and} \bibinfo{person}{Pasquale
  Malacaria}.} \bibinfo{year}{2016}\natexlab{}.
\newblock \showarticletitle{Multi-run Side-Channel Analysis Using Symbolic
  Execution and Max-SMT}. In \bibinfo{booktitle}{{\em {IEEE} 29th Computer
  Security Foundations Symposium, {CSF} 2016, Lisbon, Portugal, June 27 - July
  1, 2016}}. \bibinfo{pages}{387--400}.
\newblock


\bibitem[\protect\citeauthoryear{Pasareanu and Rungta}{Pasareanu and
  Rungta}{2010}]%
        {PasareanuR10}
\bibfield{author}{\bibinfo{person}{Corina~S. Pasareanu} {and}
  \bibinfo{person}{Neha Rungta}.} \bibinfo{year}{2010}\natexlab{}.
\newblock \showarticletitle{Symbolic PathFinder: symbolic execution of Java
  bytecode}. In \bibinfo{booktitle}{{\em {ASE} 2010, 25th {IEEE/ACM}
  International Conference on Automated Software Engineering, Antwerp, Belgium,
  September 20-24, 2010}}. \bibinfo{pages}{179--180}.
\newblock


\bibitem[\protect\citeauthoryear{Phan, Bang, Pasareanu, Malacaria, and
  Bultan}{Phan et~al\mbox{.}}{2017}]%
        {PhanBPMB17}
\bibfield{author}{\bibinfo{person}{Quoc{-}Sang Phan}, \bibinfo{person}{Lucas
  Bang}, \bibinfo{person}{Corina~S. Pasareanu}, \bibinfo{person}{Pasquale
  Malacaria}, {and} \bibinfo{person}{Tevfik Bultan}.}
  \bibinfo{year}{2017}\natexlab{}.
\newblock \showarticletitle{Synthesis of Adaptive Side-Channel Attacks}.
\newblock \bibinfo{journal}{{\em {IACR} Cryptology ePrint Archive\/}}
  \bibinfo{volume}{2017} (\bibinfo{year}{2017}), \bibinfo{pages}{401}.
\newblock


\bibitem[\protect\citeauthoryear{Prabhu, Ramalingam, and Vaswani}{Prabhu
  et~al\mbox{.}}{2010}]%
        {PrabhuRV10}
\bibfield{author}{\bibinfo{person}{Prakash Prabhu}, \bibinfo{person}{Ganesan
  Ramalingam}, {and} \bibinfo{person}{Kapil Vaswani}.}
  \bibinfo{year}{2010}\natexlab{}.
\newblock \showarticletitle{Safe programmable speculative parallelism}. In
  \bibinfo{booktitle}{{\em PLDI}}. \bibinfo{pages}{50--61}.
\newblock


\bibitem[\protect\citeauthoryear{Ramamoorthy and Li}{Ramamoorthy and
  Li}{1977}]%
        {RamamoorthyL77}
\bibfield{author}{\bibinfo{person}{C.~V. Ramamoorthy} {and}
  \bibinfo{person}{Hon~Fung Li}.} \bibinfo{year}{1977}\natexlab{}.
\newblock \showarticletitle{Pipeline Architecture}.
\newblock \bibinfo{journal}{{\em {ACM} Comput. Surv.\/}} \bibinfo{volume}{9},
  \bibinfo{number}{1} (\bibinfo{year}{1977}), \bibinfo{pages}{61--102}.
\newblock


\bibitem[\protect\citeauthoryear{Reinman and Calder}{Reinman and
  Calder}{1998}]%
        {ReinmanC98}
\bibfield{author}{\bibinfo{person}{Glenn Reinman} {and} \bibinfo{person}{Brad
  Calder}.} \bibinfo{year}{1998}\natexlab{}.
\newblock \showarticletitle{Predictive Techniques for Aggressive Load
  Speculation}. In \bibinfo{booktitle}{{\em Proceedings of the 31st Annual
  {IEEE/ACM} International Symposium on Microarchitecture, {MICRO} 31, Dallas,
  Texas, USA, November 30 - December 2, 1998}}. \bibinfo{pages}{127--137}.
\newblock


\bibitem[\protect\citeauthoryear{Sousa and Dillig}{Sousa and Dillig}{2016}]%
        {SousaD16}
\bibfield{author}{\bibinfo{person}{Marcelo Sousa} {and} \bibinfo{person}{Isil
  Dillig}.} \bibinfo{year}{2016}\natexlab{}.
\newblock \showarticletitle{Cartesian hoare logic for verifying k-safety
  properties}. In \bibinfo{booktitle}{{\em ACM SIGPLAN Conference on
  Programming Language Design and Implementation}}. \bibinfo{pages}{57--69}.
\newblock
\showDOI{%
\url{http://dx.doi.org/10.1145/2908080.2908092}}


\bibitem[\protect\citeauthoryear{Sung, Paulsen, and Wang}{Sung
  et~al\mbox{.}}{2018}]%
        {SungPW18}
\bibfield{author}{\bibinfo{person}{Chungha Sung}, \bibinfo{person}{Brandon
  Paulsen}, {and} \bibinfo{person}{Chao Wang}.}
  \bibinfo{year}{2018}\natexlab{}.
\newblock \showarticletitle{{CANAL:} a cache timing analysis framework via
  {LLVM} transformation}. In \bibinfo{booktitle}{{\em Proceedings of the 33rd
  {ACM/IEEE} International Conference on Automated Software Engineering, {ASE}
  2018, Montpellier, France, September 3-7, 2018}}. \bibinfo{pages}{904--907}.
\newblock


\bibitem[\protect\citeauthoryear{Tromer, Osvik, and Shamir}{Tromer
  et~al\mbox{.}}{2010}]%
        {TromerOS10}
\bibfield{author}{\bibinfo{person}{Eran Tromer}, \bibinfo{person}{Dag~Arne
  Osvik}, {and} \bibinfo{person}{Adi Shamir}.} \bibinfo{year}{2010}\natexlab{}.
\newblock \showarticletitle{Efficient Cache Attacks on AES, and
  Countermeasures}.
\newblock \bibinfo{journal}{{\em J. Cryptology\/}} \bibinfo{volume}{23},
  \bibinfo{number}{1} (\bibinfo{year}{2010}), \bibinfo{pages}{37--71}.
\newblock


\bibitem[\protect\citeauthoryear{Wang and Schaumont}{Wang and
  Schaumont}{2017}]%
        {WangS17}
\bibfield{author}{\bibinfo{person}{Chao Wang} {and} \bibinfo{person}{Patrick
  Schaumont}.} \bibinfo{year}{2017}\natexlab{}.
\newblock \showarticletitle{Security by compilation: an automated approach to
  comprehensive side-channel resistance}.
\newblock \bibinfo{journal}{{\em {SIGLOG} News\/}} \bibinfo{volume}{4},
  \bibinfo{number}{2} (\bibinfo{year}{2017}), \bibinfo{pages}{76--89}.
\newblock


\bibitem[\protect\citeauthoryear{Wang, Chattopadhyay, Biswas, Mitra, and
  Roychoudhury}{Wang et~al\mbox{.}}{2019b}]%
        {WangCBMR19}
\bibfield{author}{\bibinfo{person}{Guanhua Wang}, \bibinfo{person}{Sudipta
  Chattopadhyay}, \bibinfo{person}{Arnab~Kumar Biswas}, \bibinfo{person}{Tulika
  Mitra}, {and} \bibinfo{person}{Abhik Roychoudhury}.}
  \bibinfo{year}{2019}\natexlab{b}.
\newblock \showarticletitle{{KLEESPECTRE:} Detecting Information Leakage
  through Speculative Cache Attacks via Symbolic Execution}.
\newblock \bibinfo{journal}{{\em CoRR\/}}  \bibinfo{volume}{abs/1909.00647}
  (\bibinfo{year}{2019}).
\newblock


\bibitem[\protect\citeauthoryear{Wang, Sung, and Wang}{Wang
  et~al\mbox{.}}{2019c}]%
        {WangSW19}
\bibfield{author}{\bibinfo{person}{Jingbo Wang}, \bibinfo{person}{Chungha
  Sung}, {and} \bibinfo{person}{Chao Wang}.} \bibinfo{year}{2019}\natexlab{c}.
\newblock \showarticletitle{Mitigating power side channels during compilation}.
  In \bibinfo{booktitle}{{\em Proceedings of the {ACM} Joint Meeting on
  European Software Engineering Conference and Symposium on the Foundations of
  Software Engineering, {ESEC/SIGSOFT} {FSE} 2019, Tallinn, Estonia, August
  26-30, 2019}}. \bibinfo{pages}{590--601}.
\newblock


\bibitem[\protect\citeauthoryear{Wang, Bao, Liu, Wang, Zhang, and Wu}{Wang
  et~al\mbox{.}}{2019a}]%
        {WangBLWZW19}
\bibfield{author}{\bibinfo{person}{Shuai Wang}, \bibinfo{person}{Yuyan Bao},
  \bibinfo{person}{Xiao Liu}, \bibinfo{person}{Pei Wang},
  \bibinfo{person}{Danfeng Zhang}, {and} \bibinfo{person}{Dinghao Wu}.}
  \bibinfo{year}{2019}\natexlab{a}.
\newblock \showarticletitle{Identifying Cache-Based Side Channels through
  Secret-Augmented Abstract Interpretation}. In \bibinfo{booktitle}{{\em 28th
  {USENIX} Security Symposium, {USENIX} Security 2019, Santa Clara, CA, USA,
  August 14-16, 2019}}. \bibinfo{pages}{657--674}.
\newblock


\bibitem[\protect\citeauthoryear{Wang, Wang, Liu, Zhang, and Wu}{Wang
  et~al\mbox{.}}{2017}]%
        {WangWLZW17}
\bibfield{author}{\bibinfo{person}{Shuai Wang}, \bibinfo{person}{Pei Wang},
  \bibinfo{person}{Xiao Liu}, \bibinfo{person}{Danfeng Zhang}, {and}
  \bibinfo{person}{Dinghao Wu}.} \bibinfo{year}{2017}\natexlab{}.
\newblock \showarticletitle{Cache{D}: Identifying Cache-Based Timing Channels
  in Production Software}. In \bibinfo{booktitle}{{\em 26th {USENIX} Security
  Symposium ({USENIX} Security 17)}}. \bibinfo{pages}{235--252}.
\newblock


\bibitem[\protect\citeauthoryear{Weisse, Van~Bulck, Minkin, Genkin, Kasikci,
  Piessens, Silberstein, Strackx, Wenisch, and Yarom}{Weisse
  et~al\mbox{.}}{2018}]%
        {WeisseVMGKPSSWY18}
\bibfield{author}{\bibinfo{person}{Ofir Weisse}, \bibinfo{person}{Jo
  Van~Bulck}, \bibinfo{person}{Marina Minkin}, \bibinfo{person}{Daniel Genkin},
  \bibinfo{person}{Baris Kasikci}, \bibinfo{person}{Frank Piessens},
  \bibinfo{person}{Mark Silberstein}, \bibinfo{person}{Raoul Strackx},
  \bibinfo{person}{Thomas~F. Wenisch}, {and} \bibinfo{person}{Yuval Yarom}.}
  \bibinfo{year}{2018}\natexlab{}.
\newblock \showarticletitle{{Foreshadow-NG}: Breaking the Virtual Memory
  Abstraction with Transient Out-of-Order Execution}.
\newblock \bibinfo{journal}{{\em Technical report\/}} (\bibinfo{year}{2018}).
\newblock


\bibitem[\protect\citeauthoryear{Wichelmann, Moghimi, Eisenbarth, and
  Sunar}{Wichelmann et~al\mbox{.}}{2018}]%
        {WichelmannMES18}
\bibfield{author}{\bibinfo{person}{Jan Wichelmann}, \bibinfo{person}{Ahmad
  Moghimi}, \bibinfo{person}{Thomas Eisenbarth}, {and} \bibinfo{person}{Berk
  Sunar}.} \bibinfo{year}{2018}\natexlab{}.
\newblock \showarticletitle{MicroWalk: {A} Framework for Finding Side Channels
  in Binaries}. In \bibinfo{booktitle}{{\em Proceedings of the 34th Annual
  Computer Security Applications Conference, {ACSAC} 2018, San Juan, PR, USA,
  December 03-07, 2018}}. \bibinfo{pages}{161--173}.
\newblock


\bibitem[\protect\citeauthoryear{Wu, Guo, Schaumont, and Wang}{Wu
  et~al\mbox{.}}{2018}]%
        {WuGSW18}
\bibfield{author}{\bibinfo{person}{Meng Wu}, \bibinfo{person}{Shengjian Guo},
  \bibinfo{person}{Patrick Schaumont}, {and} \bibinfo{person}{Chao Wang}.}
  \bibinfo{year}{2018}\natexlab{}.
\newblock \showarticletitle{Eliminating timing side-channel leaks using program
  repair}. In \bibinfo{booktitle}{{\em Proceedings of the 27th {ACM} {SIGSOFT}
  International Symposium on Software Testing and Analysis, {ISSTA} 2018,
  Amsterdam, The Netherlands, July 16-21, 2018}}. \bibinfo{pages}{15--26}.
\newblock


\bibitem[\protect\citeauthoryear{Wu and Wang}{Wu and Wang}{2019}]%
        {WuW19}
\bibfield{author}{\bibinfo{person}{Meng Wu} {and} \bibinfo{person}{Chao Wang}.}
  \bibinfo{year}{2019}\natexlab{}.
\newblock \showarticletitle{Abstract Interpretation under Speculative
  Execution}. In \bibinfo{booktitle}{{\em ACM SIGPLAN Conference on Programming
  Language Design and Implementation}}. \bibinfo{pages}{57--69}.
\newblock


\bibitem[\protect\citeauthoryear{Yan, Fletcher, and Torrellas}{Yan
  et~al\mbox{.}}{2018}]%
        {YanFT18}
\bibfield{author}{\bibinfo{person}{Mengjia Yan},
  \bibinfo{person}{Christopher~W. Fletcher}, {and} \bibinfo{person}{Josep
  Torrellas}.} \bibinfo{year}{2018}\natexlab{}.
\newblock \showarticletitle{Cache Telepathy: Leveraging Shared Resource Attacks
  to Learn {DNN} Architectures}.
\newblock \bibinfo{journal}{{\em CoRR\/}}  \bibinfo{volume}{abs/1808.04761}
  (\bibinfo{year}{2018}).
\newblock
\showeprint[arxiv]{1808.04761}
\showURL{%
\url{http://arxiv.org/abs/1808.04761}}


\bibitem[\protect\citeauthoryear{Yarom and Falkner}{Yarom and Falkner}{2014}]%
        {YaromF14}
\bibfield{author}{\bibinfo{person}{Yuval Yarom} {and} \bibinfo{person}{Katrina
  Falkner}.} \bibinfo{year}{2014}\natexlab{}.
\newblock \showarticletitle{{FLUSH+RELOAD:} {A} High Resolution, Low Noise,
  {L3} Cache Side-Channel Attack}. In \bibinfo{booktitle}{{\em Proceedings of
  the 23rd {USENIX} Security Symposium, San Diego, CA, USA, August 20-22,
  2014.}} \bibinfo{pages}{719--732}.
\newblock


\bibitem[\protect\citeauthoryear{Zhang, Gao, Song, and Wang}{Zhang
  et~al\mbox{.}}{2018}]%
        {ZhangGSW18}
\bibfield{author}{\bibinfo{person}{Jun Zhang}, \bibinfo{person}{Pengfei Gao},
  \bibinfo{person}{Fu Song}, {and} \bibinfo{person}{Chao Wang}.}
  \bibinfo{year}{2018}\natexlab{}.
\newblock \showarticletitle{SCInfer: Refinement-Based Verification of Software
  Countermeasures Against Side-Channel Attacks}. In \bibinfo{booktitle}{{\em
  Computer Aided Verification - 30th International Conference, {CAV} 2018, Held
  as Part of the Federated Logic Conference, FloC 2018, Oxford, UK, July 14-17,
  2018, Proceedings, Part {II}}}. \bibinfo{pages}{157--177}.
\newblock


\bibitem[\protect\citeauthoryear{Zhang, Chen, and Wang}{Zhang
  et~al\mbox{.}}{2012}]%
        {ZhangCW12}
\bibfield{author}{\bibinfo{person}{Yufeng Zhang}, \bibinfo{person}{Zhenbang
  Chen}, {and} \bibinfo{person}{Ji Wang}.} \bibinfo{year}{2012}\natexlab{}.
\newblock \showarticletitle{Speculative Symbolic Execution}. In
  \bibinfo{booktitle}{{\em 23rd {IEEE} International Symposium on Software
  Reliability Engineering, {ISSRE} 2012, Dallas, TX, USA, November 27-30,
  2012}}. \bibinfo{pages}{101--110}.
\newblock


\end{thebibliography}
